\documentclass[%
 amsmath,amssymb,
 aps, physrev,twocolumn,
]{revtex4-2}

\usepackage{graphicx}
\usepackage{dcolumn}
\usepackage{bm}
\usepackage{float}
\newcommand{\ket}[1]{\left|#1\right\rangle}

\usepackage{xcolor}

\begin{document}

\preprint{APS/123-QED}

\title{Elastic and Quasielastic Electron Scattering in Perfect Crystals}

 \author{Eric J.~Heller} 
\affiliation{Department of Physics, Harvard University, Cambridge, MA 02138, USA}
\affiliation{Department of Chemistry and Chemical Biology, Harvard University, Cambridge,
MA 02138, USA}
\author{Anton M.~Graf}
\affiliation{Harvard John A. Paulson School of Engineering and Applied Sciences,
Harvard, Cambridge, Massachusetts 02138, USA}
\affiliation{Department of Chemistry and Chemical Biology, Harvard University, Cambridge,
MA 02138, USA}
\author{Yubo Zhang}\affiliation{Quantum Science and Engineering, Harvard University, Cambridge, MA 02138, USA}

\author{Alhun~Aydin}
\affiliation{Faculty of Engineering and Natural Sciences, Sabanci University, 34956 Tuzla, Istanbul, Turkey}
\affiliation{Department of Physics, Harvard University, Cambridge, MA 02138, USA}

 \author{Joonas~Keski-Rahkonen}
\affiliation{Department of Physics, Harvard University, Cambridge, MA 02138, USA}

\begin{abstract}
Momentum relaxation of electrons in perfect crystals is conventionally
associated with phonon emission or absorption and is therefore regarded
as intrinsically inelastic. We show that this is not a fundamental
requirement. Starting from the Fr\"ohlich Hamiltonian, but making fewer
assumptions than usual—retaining total mechanical momentum and the full
density-density electron-lattice interaction—we find a substantially
larger class of allowed processes. When total momentum conservation is enforced together with all
pseudomomenta, the lattice center-of-mass provides a continuous momentum
reservoir, and phonons are no longer required to account for the
electron’s momentum change. The conventional phonon-mediated processes
emerge as a restricted subset of a higher-dimensional manifold of
transitions. Within this expanded space, elastic and quasielastic momentum-transfer
channels appear already at first order, even in a perfect crystal.
Momentum relaxation can therefore proceed rapidly while energy relaxation
remains slow, consistent with a wide range of experimental observations
including hot-electron injection, quantum oscillations, weak localization,
and Coulomb drag in low-dimensional conductors. In particular, scattering
in a perfect one-dimensional wire is not kinematically constrained once
the lattice recoil degree of freedom is retained.

\end{abstract}

\maketitle

\section{Introduction}

Electron–lattice interactions play a central role in condensed matter physics.
They govern both momentum relaxation and energy exchange in crystalline solids,
and therefore underlie electrical resistivity and the loss of electronic coherence.

In the modern transport theory of metals (Bloch–Peierls–Ziman–Mott–Ashcroft–Mermin), the electron–lattice problem is formulated by imposing periodic boundary conditions and expanding lattice displacements using internal normal modes. In this construction, going back to Born and von K'arm'an, the lattice center-of-mass (zero) mode is removed at the outset, leaving only internal vibrational degrees of freedom. When an electron applies force to a lattice atom, it can only respond with finite energy phonons.  The zero energy recoil, without which M\"ossbauer scattering would falter, has been removed. In first order, momentum transfer is  forced to reside in a single phonon, leading to the familiar selection rules $k’ = k \pm q$ and $\varepsilon_{k’} = \varepsilon_k \pm \hbar\omega_q$. In a perfect crystal, a change in electron momentum thus requires phonon emission or absorption and is therefore inelastic. Although this is true within the formulation, it is an unnatural result, not followed in the laboratory. 


 When the lattice center-of-mass degree of freedom is
retained, the total pseudomomentum is conserved and can be exchanged
between the electron and the background. If the zero mode is removed,
this exchange channel is absent. The phonon sector alone does not form a
closed system for pseudomomentum, and the familiar selection rule
\[
\Delta K_e+\Delta K_{\rm ph}=Gm
\]
need not hold for individual processes. Indeed, it is violated once the
density-density interaction is treated beyond the linear
Fr\"ohlich approximation. Pseudomomentum is therefore not a conserved
quantity within the electron-phonon subspace alone. The traditional clean pseudomomentum accounting falters if even the next terms are kept in the the density-density expansion, e.g.

Purely elastic processes are a rarity in the classical world, but are commonplace in the quantum world. Consider a linear atom-diatom collision, i.e. three translational degrees of freedom, separating to a center mass of the diatom, the atom, and the internal virbatipnal coordinate: if not too violent, upon measurement the diatom is likely to be in the state it started in, even though it has definitely been disturbed. Again, M\"ossbauer emission comes to mind. In fact for pass-through experiments, like fast electrons, both Bragg and a good portion of diffuse scattering is known to be elastic, with the crystal recoiling as a whole. Somehow an electron residing inside a crystal has found itself treated a different way, but it should not be.

Experimentally,  momentum and energy relaxation often occur on
markedly different timescales.
In clean metals over broad temperature ranges, electrical resistivity and
magnetotransport measurements indicate rapid momentum randomization,
while hot-electron experiments, Johnson-noise thermometry, and related
nonequilibrium probes show that energy relaxation to the lattice can remain
comparatively slow.
The resulting hierarchy
\begin{equation}
\tau_p \ll \tau_E ,
\end{equation}
where $\tau_p$ is the momentum-relaxation time and $\tau_E$ the
energy-relaxation time, indicates that electronic momentum can be efficiently
scrambled without substantial net energy transfer to internal lattice excitations.

The experimentally observed separation between $\tau_p$ and $\tau_E$ does not seem to comport with the  ``always inelastic'' narrative, and this reality check motivates a
careful examination of how momentum conservation is implemented
in a perfectly periodic crystal.

At finite temperature, lattice vibrations introduce perturbations that
allow electrons to scatter between Bloch states. This mechanism was first
analyzed by Bloch, who obtained the characteristic low-temperature
$T^{5}$ resistivity associated with electron-phonon scattering in metals
\cite{Bloch1930}. These ideas were incorporated into the
Bloch-Gr\"uneisen-Boltzmann transport framework
\cite{Bloch1930,Gruneisen1933}, in which the electronic distribution
function evolves according to a Boltzmann equation for Bloch electrons.
Within this formulation, scattering is described as transitions between
Bloch states accompanied by phonon absorption or emission, enforced by
energy-conserving delta functions. In the standard Fr\"ohlich
Hamiltonian~\cite{Frohlich1954}, the interaction is expressed in terms of
phonon creation and annihilation operators, so that momentum transfer is
associated, at lowest order, with changes in phonon occupation and is
therefore intrinsically inelastic, subject to pseudomomentum and energy
conservation \cite{Ziman,AshcroftMermin}. Elastic momentum relaxation is
accordingly attributed to impurities or other forms of explicit disorder.
The subsequent development of Migdal-Eliashberg theory further
reinforced this picture, embedding the identification of momentum
relaxation with inelastic electron-phonon scattering into the standard
theoretical framework.

The situation has been different for external, ``pass through'' probes such as X-rays or neutrons or even fast electrons. 
In such circumstances, elastic and even diffuse elastic recoil of the entire crystal is routinely included through
Debye-Waller theory. Here, we will see that there is no solid basis for the asymmetry in the treatment of an interior electron.



As we will see, bringing back the center of mass dramatically changes the narrative of internal scattering processes. Somehow that seems implausible, which is one reason perhaps that the inelastic narrative has persisted for so long.  

We revisit this issue determined to be rigorous
about momentum conservation in a fully isolated electron-crystal system with all applicable momenta retained. (The isolation can later be broken once a full understanding is reached of the isolated system.)
When  the electron-lattice interaction
is written in its full density-density form, elastic and quasi-elastic
momentum exchange channels appear already at first order, without any defects present other than the transient ones implied by the thermal deformation potential. 

A few preliminary, qualitative remarks are in order.  First, removing the center of mass makes electrons in crystals is akin to the venerable particle in a box model. There too there is no center of mass of the box, and no box recoil. Momentum is not conserved.  In the Appendix we show what happens when the box is isolated, given  mass, and allowed to recoil. Second, if the center of mass is absent, the only degree of freedom that could carry the recoil resulting from an elastic interior electron deflection is also absent. Phonons will be forced to take on that role and indeed \textit{by construction} there can be no elastic events of this sort.



 If the crystal  total mechanical
momentum is retained, and total momentum conservation is again enforceable,
a transfer of momentum $\mathbf q$ to an electron is accompanied by an
equal and opposite momentum $-\mathbf q$ carried by the remaining
degrees of freedom, which we identify as the lattice background. Those degrees of freedom include the several lattice zero frequency recoil modes, and internal vibrational modes. 





An instructive analog is provided by M\"ossbauer scattering, in which
the dominant elastic line, perhaps 97\% of the total probability, implies recoil of the entire lattice
without internal phonon excitation. 


We hasten to point out that the question here involves real physics, and not mere representation: are phonons   created or destroyed when electrons deflect in a pure crystal, or not?  That is an essential, physical difference, with profound experimental consequences. 




In a finite-temperature many-body system, strictly elastic processes
do not exist in a  provable, absolute sense.
Even the dominant ``elastic'' line in M\"ossbauer scattering exhibits slight
thermal broadening.
We therefore use the term \emph{elastic} to denote processes in which
phonon occupation does not change, and \emph{quasielastic} to denote
processes involving small energy exchanges that do not balance electron
momentum transfer; that task will be left to the crystal center of mass recoil. In fact there are so many ways that momentum conservation can be accommodated that the textbook phonon carrying away exactly the right momentum is a measure zero event, in spite of its utility as a model. This will become clearer below.

\section{A minimal example: the electron on a free atomic ring}
\label{ring}
\begin{figure*}[t]
    \centering
\includegraphics[width=0.85\linewidth]{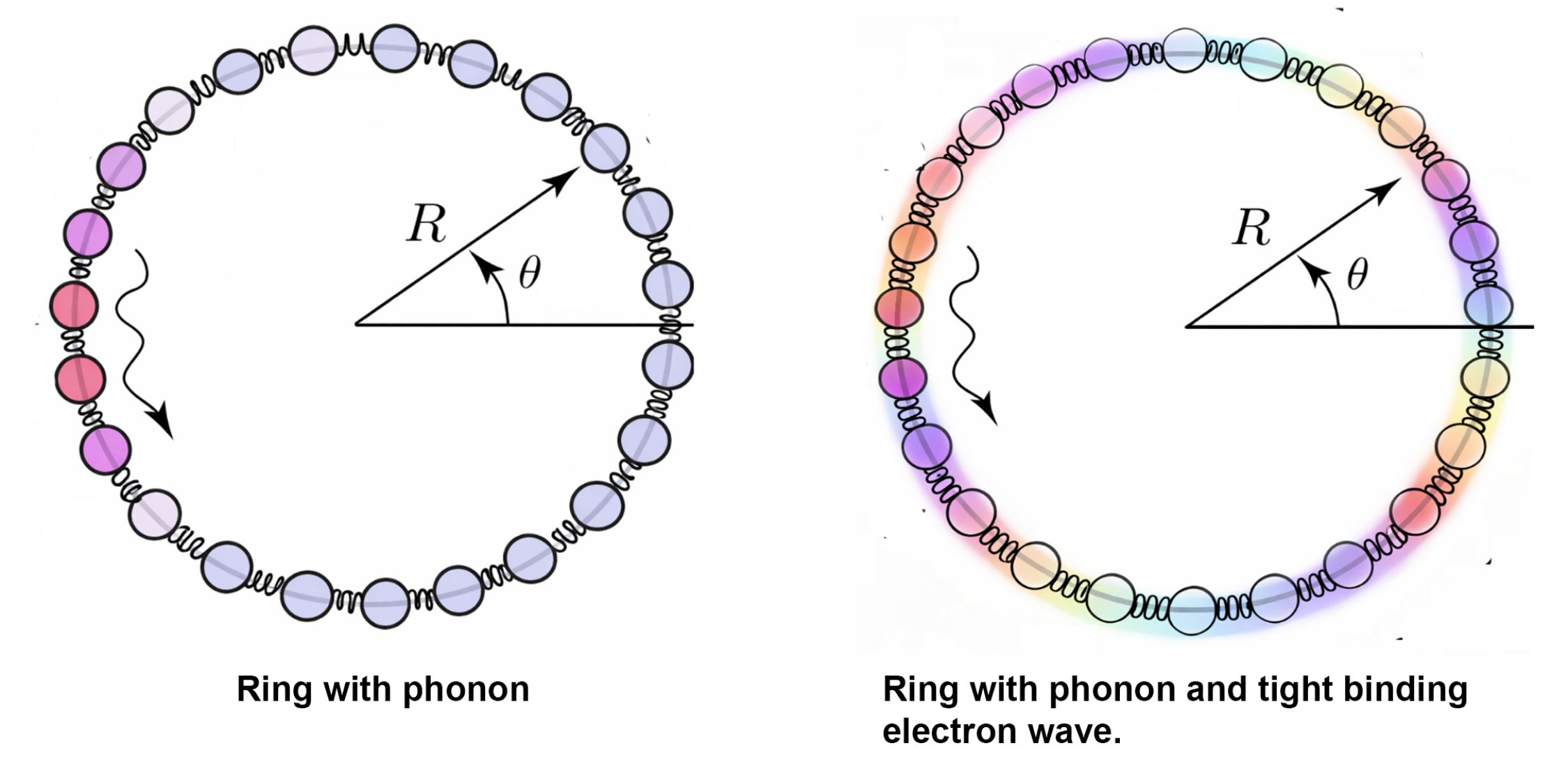}
    \caption{
\textbf{Left:} Ring lattice with a localized phonon excitation. All angular coordinates are retained, including the 0 mode (angular momentum). 
\textbf{Right:} The tight-binding electronic wavefunction is added.
The color wheel represents the amplitude and phase of the electronic amplitude
$e^{ikx}$, winding here by $4\pi$ around the ring. Only total angular momentum $L_{\rm tot}=L+R\hbar k$.
 and energy are conserved quantities.
}
    \label{fig:rings}
\end{figure*}
Before turning to the general many-body problem, it is useful to display
the dramatic effect of including the zero mode  in the simplest finite crystal that has exact momentum
bookkeeping: a ring of $N$ identical atoms of mass $m$, constrained to
move on a circle of radius $R$ and coupled by nearest-neighbor springs (figure~\ref{fig:rings}, left). Completing the picture,
an electron moves on the same ring by {\it dynamic} tight binding nearest-neighbor hopping (figure~\ref{fig:rings}, right). Dynamic means the atoms   move along the ring, unconstrained except for their neighbors, and carry their orbitals with them. An early example of this approach applied to graphene is found in ref.~\cite{mohanty_lazy_2019}.
The importance of this model is its simplicity, yet it retains essential  electron-phonon interactions. The lattice
zero mode is present explicitly, total angular momentum is conserved,
and the distinction between internal phonons and rigid-body recoil is
completely transparent. All lattice degrees of freedom are retained, including the global rotational (zero) mode, which carries the total angular momentum of the lattice.

    The standard conclusion for such a 1D model, like a perfect wire, is
    that only elastic backscattering is allowed, and even then under
    restricted circumstances, due to the simultaneous constraints of energy
    and momentum conservation when only the electron and a phonon are
    present. Efficient relaxation requires an additional degree of
    freedom. In conventional treatments\cite{Lunde2007ThreeParticle,Karzig2010HotElectrons} this is introduced in the form of a
    third electronic excitation, such as a deep hole drawn from the Fermi
    sea to satisfy the combined energy and momentum
    constraints that cannot be met within a strictly two-body
    electron-phonon process. Multiphonon processes may also be invoked, but  at
    higher order.  
    An especially clean case is provided by Coulomb drag\cite{Levchenko2011InteractionCorrections} in wire pairs, where exponential suppression of drag with temperature is again predicted, but instead power law dependence is seen in  experiments.  We will find something completely different, with manifolds of transitions possible obeying all conserved quantities. The third, ``new'' degree of freedom turns out to be the long lost center of mass.

    In drag, current in one wire creates fluctuating electric fields that transfer momentum to the other wire through the Coulomb interaction. For that transfer to occur, both wires must be able to absorb the same small energy $\hbar\omega$ and momentum $q$. The drag rate is therefore governed by the overlap of the two wires’ low-energy dynamical response functions. If each wire has gapless excitations near the Fermi points, then as $T\to 0$ there remains a shrinking but nonzero window of allowed $\omega$ and $q$, typically of size set by $k_B T$. Integrating over that thermal window gives a power of $T$.

\subsection{Hamiltonian and collective coordinates}

Let $\phi$ denote the angular coordinate of the electron, and let
$\theta_\ell$ denote the instantaneous angular position of atom
$\ell=1,\dots,N$ on the ring. The atomic coordinates are dynamical
variables and include thermal displacements.

For each instantaneous lattice configuration $\{\theta_\ell\}$, define a
set of localized electronic orbitals
\begin{equation}
\chi_\ell(\phi;\theta_\ell)=\chi(\phi-\theta_\ell),
\end{equation}
where $\chi(\phi-\theta_\ell)$ is an electron wavefunction centered on
the nucleus at $\theta_\ell$. A general single-electron state may then
be written as
\begin{equation}
\Psi(\phi;\{\theta_\ell\})
=
\sum_{\ell=1}^N
\psi_\ell\,\chi(\phi-\theta_\ell),
\label{eq:lattice_following_expansion}
\end{equation}
where the coefficients $\psi_\ell$ give the electronic amplitudes on
the lattice-following orbitals.

The full Hamiltonian acts both on the lattice coordinates
$\{\theta_\ell\}$ and on the electronic amplitudes $\{\psi_\ell\}$:
\begin{equation}
H
=
H_{\rm lat}(\{\theta_\ell\})
+
H_{\rm el}(\{\theta_\ell\}),
\label{eq:full_H_tb}
\end{equation}
Here $H_{\rm el}(\{\theta_\ell\})$ denotes the electronic Hamiltonian
in the lattice-following orbital basis. It encodes the electron-lattice interaction. Its matrix elements depend on
the instantaneous atomic positions and therefore include both the
electronic energies and the electron-lattice interaction.
\begin{equation}
\begin{aligned}
H_{\rm lat}
&=
-\frac{\hbar^2}{2I_{\rm ion}}
\sum_{\ell=1}^N
\frac{\partial^2}{\partial \theta_\ell^2}
+
\frac{K}{2}\sum_{\ell=1}^N
\left(\theta_{\ell+1}-\theta_\ell-\tfrac{2\pi}{N}\right)^2 ,
\\
&\qquad
\theta_{N+1}\equiv \theta_1+2\pi .
\end{aligned}
\label{eq:Hlat_theta}
\end{equation}

In the nearest-neighbor tight-binding approximation, the electronic
Hamiltonian acts on the coefficients $\psi_\ell$ as
\begin{equation}
(H_e\psi)_\ell
=
-\,t_{\ell-1}\,\psi_{\ell-1}
-\,t_\ell\,\psi_{\ell+1},
\qquad
\psi_{\ell+N}\equiv \psi_\ell ,
\label{eq:He_amplitudes}
\end{equation}
where the hopping amplitudes $t_\ell$ are determined by the overlap of
neighboring orbitals and therefore depend on the instantaneous bond
lengths, i.e. on the differences $\theta_{\ell+1}-\theta_\ell$.
\subsection{Lattice: zero mode and internal phonons}

Let $\theta_\ell$ be the angular position of atom $\ell=1,\dots,N$, with
periodic indexing $\theta_{N+1}\equiv\theta_1$.  Introduce the center angle
\begin{equation}
\Theta=\frac{1}{N}\sum_{\ell=1}^N \theta_\ell
\end{equation}
and internal coordinates
\begin{equation}
\varphi_\ell=\theta_\ell-\Theta,
\qquad
\sum_{\ell=1}^N \varphi_\ell=0 .
\end{equation}
The angular coordinates $\theta_\ell$ denote the instantaneous positions
of the atoms and include thermal displacements. We therefore write
\begin{equation}
\theta_\ell = \Theta + \frac{2\pi \ell}{N} + \varphi_\ell,
\end{equation}
where $\Theta$ describes the rigid rotation of the lattice and
$\varphi_\ell$ the internal (thermal) displacements, with
$\sum_\ell \varphi_\ell = 0$.

The lattice Hamiltonian is
\begin{equation}
H_{\rm lat}
=
\sum_{\ell=1}^N \frac{L_\ell^2}{2mR^2}
+
\frac{\kappa R^2}{2}
\sum_{\ell=1}^N (\varphi_{\ell+1}-\varphi_\ell)^2 .
\end{equation}
In these variables the Hamiltonian separates exactly as
\begin{equation}
H_{\rm lat}
=
\frac{L^2}{2I_{\rm tot}}
+
H_{\rm ph},
\qquad
I_{\rm tot}=NmR^2,
\label{eq:ring_sep}
\end{equation}
where
\begin{equation}
L=\sum_{\ell=1}^N L_\ell
=
-i\hbar\frac{\partial}{\partial\Theta}
\end{equation}
is the angular momentum of rigid rotation of the entire ring.  
The term $H_{\rm ph}$ contains the $N-1$ internal normal modes built from
the coordinates $\varphi_\ell$.

Thus the $q=0$ mode is not discarded: it appears explicitly as the rotor
term $L^2/2I_{\rm tot}$, while the remaining modes are the usual phonons.

In the lattice-following description, the electron is represented in a
basis of localized orbitals $\chi(\phi-\theta_\ell)$ centered on the
instantaneous atomic positions. A general state may be written as
\begin{equation}
\Psi(\phi;\{\theta_\ell\})
=
\sum_{\ell} \psi_\ell\, \chi(\phi-\theta_\ell),
\end{equation}
where $\psi_\ell$ are the amplitudes on the orbitals.

In this basis the electronic Hamiltonian acts on the amplitudes through
nearest-neighbor matrix elements,
\begin{equation}
(H_{\rm el}\psi)_\ell
=
\varepsilon_\ell(\{\theta\})\,\psi_\ell
-
t_{\ell-1}(\{\theta\})\,\psi_{\ell-1}
-
t_\ell(\{\theta\})\,\psi_{\ell+1},
\end{equation}
with periodic boundary conditions $\psi_{\ell+N}\equiv\psi_\ell$.
The matrix elements depend on the instantaneous lattice configuration
and therefore incorporate both the electronic structure and the
electron-lattice coupling.

For small distortions, the bond length depends only on internal
differences,
\begin{equation}
d_\ell \simeq a + R(\varphi_{\ell+1}-\varphi_\ell),
\qquad
a=\frac{2\pi R}{N},
\end{equation}
so that, to linear order,
\begin{equation}
t_\ell \simeq t_0 - \alpha R(\varphi_{\ell+1}-\varphi_\ell).
\end{equation}
Thus the electronic Hamiltonian depends only on the internal
coordinates $\{\varphi_\ell\}$: the rigid rotation $\Theta$ drops out.

\subsection{Relative coordinate and Bloch-like structure}

Because the Hamiltonian depends on $\phi$ and $\Theta$ only through the
relative coordinate $\chi=\phi-\Theta$, the eigenstates may be chosen in
Bloch-like form,
\begin{equation}
\Psi_k(\phi,\Theta,\{\varphi_\ell\})
=
e^{ik(\phi-\Theta)}
\sum_{\ell}
u_{\ell k}(\{\varphi_\ell\})\,
\chi(\phi-\theta_\ell),
\label{eq:diabatic_ansatz}
\end{equation}
where the coefficients $u_{\ell k}$ describe the internal structure of
the state in the lattice-following basis.

Here the instantaneous atomic positions are
\begin{equation}
\theta_\ell
=
\Theta + \frac{2\pi \ell}{N} + \varphi_\ell,
\end{equation}
so that the orbitals $\chi(\phi-\theta_\ell)$ move with the lattice atoms11.

In this form the exponential factor $e^{ik(\phi-\Theta)}$ carries the
crystal pseudomomentum, while the coefficients $u_{\ell k}$ encode the
internal, lattice-following structure of the electronic state. In the
absence of distortions,
$u_{\ell k} \propto e^{ik \frac{2\pi \ell}{N}}$, and
Eq.~\eqref{eq:diabatic_ansatz} reduces to the standard tight-binding
Bloch wave.

The factor $e^{ik(\phi-\Theta)}$ in equation~\ref{eq:diabatic_ansatz} is revealing and promising. The overall zero mode rotation angle $\Theta$ now takes its place among the dynamical variables. It is not benign, because it is not quite the angle belonging to the the total angular momentum, since that includes the electron. The key point is that the evolution of $\Theta$ is nontrivial, depending on recoil from changes in the angular momentum of the electron. It is clear from this form that if the electron $\phi$ has pseudomomentum, so must $\Theta$, and   both can change, always equally and oppositely. This will be developed in the next section. The sum of the two pseudomomenta, call them $k_\phi =k$ and $k_\Theta=-k$, vanishes, and remains zero under changes of $k$, assuring the the conservation of total pseudomomentum. 

Because zero mode rotation it is now included, a  world of elastic and quasielastic processess will open up.

\subsection{Total angular momentum balance}

Rotational invariance of the full Hamiltonian implies conservation of
the total angular momentum operator
\begin{equation}
L_{\rm tot}
=
-i\hbar\frac{\partial}{\partial\phi}
-
i\hbar\frac{\partial}{\partial\Theta}.
\end{equation}

Acting on the Bloch-like state gives the one-line conservation rule
\begin{equation}
L_{\rm tot}
=
L + R\hbar k ,
\label{eq:one_line_conservation}
\end{equation}
where $L$ is the angular momentum conjugate to $\Theta$.

Equation~\eqref{eq:one_line_conservation} implies that any change in the
electronic pseudomomentum $k$ must be accompanied by a compensating
change in $L$, corresponding to recoil of the ring as a whole. This is of course necessary to keep total  angular momentum constant (lattice plus electron).  Interestingly, there is no role yet for phonon momentum, and we can leave it this way if we want. As Peierls pointed out, phonon momentum can be assigned to the lattice as a whole\cite{Peierls1955}, Chpts 1-2. Here the momentum appeared as belonging to  the lattice, and indeed in must, since $L=\sum_\ell L_\ell,$ which is   a sum over all the motion of every atom. Any supposed phonon momentum would be double counting. 

We have already remarked that the form $\exp[ik(\psi-\Theta)] $ already implies the $L$ (not $L_{tot}$) must be regarded as a pseudomomentum.  This will br defended on symmetry grounds in the next section.  Both the electron and its host have pseudomomentum, which is actually a very comfortable circumstance. We should indeed be uncomfortable if we suppose that a pseudomomentum and a mechanical momentum could combine. 




\subsection{Phonon-diagonal (elastic) backscattering}

The simplest illustration of the dramatic effect of keeping the zero recoil mode is provided by
phonon-diagonal backscattering. In the absence of lattice distortions,
the electronic eigenstates are plane waves
\begin{equation}
\ket{k}=\frac{1}{\sqrt N}\sum_{\ell} e^{ika\ell} c_\ell^\dagger\ket{0},
\end{equation}
with even dispersion
$
\varepsilon(k)=\varepsilon(-k).$
For an initial state
\begin{equation}
\ket{\Psi_i}=\ket{k}\ket{\nu}\ket{L},
\end{equation}
one allowed elastic transition is
\begin{equation}
\ket{k}\ket{\nu}\ket{L}
\;\rightarrow\;
\ket{-k}\ket{\nu}\ket{L+2\hbar kR},
\label{eq:ring_backscatter}
\end{equation}
in which the electron reverses its direction while the internal phonon
state remains unchanged. The compensating angular momentum is absorbed
by rigid-body recoil of the lattice.

The elastic backscattering channel
\[
\ket{k}\ket{\nu}\to \ket{-k}\ket{\nu}
\]
is kinematically forbidden in the conventional phonon-only
representation unless Umklapp is invoked. In this process
$\Delta K_e=-2k$ while $\Delta K_{\rm ph}=0$, so the standard selection
rule cannot be satisfied by phonons alone. More fundamentally, the
process has no carrier of the corresponding mechanical momentum once the
lattice center-of-mass degree of freedom is removed. The absence of this
channel is therefore not a dynamical result, but a consequence of having
eliminated the recoil degree of freedom that would otherwise carry the
compensating momentum.With the pseudoangularmomentum intact, it is kinematically robust,  but to  verify that this channel is dynamically allowed, we evaluate the
corresponding matrix element of the electron-lattice interaction. In
the density-density (bond-length dependent) form, the hopping amplitude
depends on the relative displacements of neighboring sites,
\(
t_\ell = t_0 + \alpha (u_{\ell+1}-u_\ell).
\)
Because the interaction depends only on internal coordinate differences,
it is independent of the global rotational coordinate, and therefore does
not couple directly to $L$.

\paragraph*{Nonvanishing elastic recoil matrix element (moving basis).}

We evaluate the matrix element in an instantaneous lattice basis, in
which localized electronic orbitals follow the atomic positions.  Let
\begin{equation}
R_\ell(\Theta,\{\xi\}) = R\Theta + \ell a + \xi_\ell ,
\end{equation}
where $\Theta$ is the global rotational coordinate and $\{\xi_\ell\}$
are internal displacements. The electronic states are
\begin{equation}
\ket{k;\Theta,\{\xi\}}
=
\frac{1}{\sqrt N}\sum_{\ell} e^{ik\ell a}\,
c_\ell^\dagger(\Theta,\{\xi\})\ket{0}.
\end{equation}

For fixed configuration, the Hamiltonian is
\begin{equation}
H_e =
-\sum_{\ell} t_\ell(\{\xi\})
\left(c_{\ell+1}^\dagger c_\ell + \mathrm{h.c.}\right),
\end{equation}
with
\begin{equation}
t_\ell(\{\xi\}) = t\!\left(a+\xi_{\ell+1}-\xi_\ell\right).
\end{equation}
The global coordinate $\Theta$ drops out, as $t_\ell$ depends only on
internal differences.

For the elastic recoil channel
\begin{equation}
\ket{k,\nu,L}
\rightarrow
\ket{-k,\nu,L+2\hbar kR},
\end{equation}
the electronic matrix element at fixed configuration is
\begin{equation}
\langle -k|H_e|k\rangle
=
-\frac{2\cos ka}{N}
\sum_{\ell} t_\ell(\{\xi\})\,e^{2ika\ell}.
\end{equation}
For a uniform lattice this vanishes, but for a generic instantaneous
configuration $t_\ell$ is nonuniform and the $2k$ component is
nonzero. To leading order,
\begin{equation}
t_\ell \simeq t_0 + \alpha(\xi_{\ell+1}-\xi_\ell),
\end{equation}
so that
\begin{equation}
\langle -k|H_e|k\rangle
\propto
\sum_{\ell} e^{2ika\ell}(\xi_{\ell+1}-\xi_\ell).
\end{equation}

This is a satisfying outcome: the backscattering amplitude is controlled by the $2k$ Fourier
component of the instantaneous deformation field and is generically
nonvanishing. The operator acts only on internal coordinates, while the
required compensating change is carried by the global mode,
$L\to L+2\hbar kR$. The process therefore remains elastic in the phonon
sector..

The associated rotational energy shift
\begin{equation}
\Delta E_{\rm rot}
=
\frac{(L+2\hbar kR)^2-L^2}{2I_{\rm tot}}
\end{equation}
is negligible for macroscopic
$I_{\rm tot}=NmR^2$, so the process is effectively elastic.

This example shows explicitly that momentum transfer need not be tied to
a change in the internal phonon state: it may instead be carried by the
lattice zero mode.

It is worth emphasizing that the phonon-diagonal backscattering process
in Eq.~\eqref{eq:ring_backscatter} is absent in standard treatments in
which the lattice center-of-mass (zero mode) is eliminated. In that
representation, momentum conservation is enforced solely through the
internal phonon modes, implying $k' = k \pm q$ and excluding the
channel $\nu'=\nu$ for $k\to -k$. The absence of this process therefore
reflects the removal of the recoil degree of freedom rather than a
dynamical constraint.
\subsection{Elastic and quasi-elastic scattering}

The general kinematics of electron-lattice scattering in the ring
geometry follow directly from the conservation law
Eq.~\eqref{eq:one_line_conservation}. If the electron changes crystal
momentum
\begin{equation}
k \rightarrow k',
\end{equation}
the lattice angular momentum must adjust according to
\begin{equation}
L' = L + \hbar R (k-k').
\label{eq:Lshift}
\end{equation}

Thus the exact momentum-conserving transitions take the form
\begin{equation}
\ket{k,\nu,L}
\;\rightarrow\;
\ket{k',\nu',L'}.
\label{eq:general_scattering}
\end{equation}

The corresponding energy balance is
\begin{equation}
\varepsilon(k)-\varepsilon(k')
=
\sum_q \hbar\omega_q (n_q'-n_q)
+
\frac{(L')^2-L^2}{2I_{\rm tot}},
\label{eq:energy_balance_full}
\end{equation}
where the final term represents the recoil energy of the lattice. For a
macroscopic ring this contribution is typically negligible, so that
\begin{equation}
\varepsilon(k)-\varepsilon(k')
\simeq
\sum_q \hbar\omega_q (n_q'-n_q).
\label{eq:energy_balance_recoil}
\end{equation}

Because the lattice angular momentum $L$ is retained as a dynamical
degree of freedom, each electronic transition $k\to k'$ is compatible
with a broad family of vibrational responses:

\begin{enumerate}

\item \textbf{Elastic recoil}
\begin{equation}
\ket{k,\nu,L}
\rightarrow
\ket{k',\nu,L+\hbar R (k-k')},
\end{equation}
in which the phonon configuration is unchanged and the rigid rotation
of the lattice carries the momentum transfer.

\item \textbf{Quasi-elastic single-phonon processes}
\begin{equation}
\ket{k,\nu,L}
\rightarrow
\ket{k',\nu\pm 1_q,L+\hbar R (k-k')},
\end{equation}
in which the zero mode absorbs the angular momentum while a phonon
adjusts the energy balance.

\item \textbf{Multiphonon processes}
\begin{equation}
\ket{k,\nu,L}
\rightarrow
\ket{k',\nu+\{\Delta n_q\},L+\hbar R (k-k')}.
\end{equation}

\item \textbf{Hybrid recoil-phonon processes}
in which the momentum transfer is shared between the rigid lattice
rotation and the internal phonon manifold.

\end{enumerate}

Restoring the zero mode therefore enlarges the accessible final states
from $\ket{k',\nu'}$ to $\ket{k',\nu',L'}$. The additional angular
momentum degree of freedom opens a wide manifold of momentum-conserving
elastic and quasi-elastic scattering channels. The interaction matrix
elements
\begin{equation}
\langle k',\nu',L' \vert H \vert k,\nu,L \rangle
\end{equation}
are therefore generically nonzero across this enlarged space of states.

\subsection{Consequences for one-dimensional transport and ring geometries}

In strictly one-dimensional electronic systems, two-particle collisions
are strongly constrained by the simultaneous conservation of energy and
momentum,
\begin{equation}
k_1+k_2 = k_1' + k_2',
\qquad
\varepsilon(k_1)+\varepsilon(k_2)
=
\varepsilon(k_1')+\varepsilon(k_2').
\end{equation}
For a monotone dispersion these relations admit only trivial
rearrangements, so that two-body processes cannot efficiently relax a
near-Fermi distribution. In particular, within conventional
electron-phonon treatments, even backscattering is highly constrained,
since momentum transfer must be accompanied by phonon creation or
annihilation, and the combined conservation laws severely restrict the
available phase space.

Standard approaches therefore invoke an additional dynamical
participant. This is typically taken to be a third electronic excitation,
\begin{equation}
k_1 + k_2 + k_3 = k_1' + k_2' + k_3',
\end{equation}
with $k_3$ a deep hole drawn from the Fermi sea
\cite{Lunde2007ThreeParticle,Karzig2010HotElectrons,Levchenko2011InteractionCorrections}.
Because such excitations are thermally rare, the resulting relaxation
rates are predicted to be exponentially suppressed at low temperature.
Multiphonon processes can also relax the constraints, but only at higher
order.

A particularly clear manifestation of this reasoning appears in Coulomb
drag between parallel quantum wires, where theory predicts exponential
suppression of drag at low temperature, in contrast with the power-law
behavior observed experimentally.

Once the lattice center-of-mass mode is retained, however, an additional
dynamical participant is always present. The conserved momentum becomes
\begin{equation}
P_{\rm tot} = P_{\rm CM} + \sum_i \hbar k_i ,
\end{equation}
or in the ring geometry,
\begin{equation}
L_{\rm tot} = L + \sum_i \hbar k_i R .
\end{equation}
An electronic momentum change $k \rightarrow k'$ can then be balanced by
recoil of the lattice,
\begin{equation}
\Delta P_{\rm CM} = -\hbar (k' - k),
\end{equation}
with associated energy
\begin{equation}
\Delta E_{\rm CM}
=
\frac{(\Delta P_{\rm CM})^2}{2M_{\rm tot}},
\end{equation}
which is negligible for macroscopic $M_{\rm tot}$.

The additional degree of freedom required by one-dimensional kinematics
is therefore always available and need not be supplied by a thermally
activated deep electronic excitation. The ``third particle'' is the
collective zero mode of the lattice.

As a result, large manifolds of elastic and quasi-elastic transitions
exist that satisfy all conservation laws. Energy redistribution occurs
within the phonon manifold rather than through rare electronic
excitations, leading naturally to power-law, rather than exponentially
suppressed, low-temperature behavior.

Such power-law relaxation is widely observed in clean quantum wires and
related one-dimensional conductors
\cite{Barak2010HotElectron,Altimiras2010Relaxation}. In this picture the
lattice zero mode provides the natural third participant in the
scattering process, enabling momentum relaxation through a broad class
of elastic and quasi-elastic channels.

The same kinematics appears in the ring geometry familiar from the
theory of persistent currents. Once the lattice center-of-mass motion is
retained, the electronic phase winding can transfer angular momentum to
the lattice through recoil, with the zero mode acting as the collective
degree of freedom that accommodates the momentum balance.

Finally, we emphasize that the appearance of this additional degree of
freedom is not an assumption but a consequence of enforcing the exact
symmetries of the underlying Hamiltonian. No new interaction has been
introduced, and no approximation has been invoked beyond those already
implicit in the microscopic model. Retaining the lattice center-of-mass
simply restores the full momentum conservation law of the isolated
system. 

Any objection that the zero mode should not act as the required third
participant must therefore confront the fact that, once this conservation
law is imposed, a large manifold of elastic and quasi-elastic scattering
channels follows directly. These channels satisfy all conservation laws
and arise without fine tuning or higher-order processes. The resulting
kinematics is thus not a hypothesis but an unavoidable consequence of the
complete Hamiltonian description.

\section{Foreground, Background, and Pseudomomentum}
\label{ForBac}




  \subsection{Defining Pseudomomentum}


It is important that we clearly distinguish component parts We distinguish   \emph{foreground} and \emph{background}
degrees of freedom.
The foreground is the object of interest, here a single electron, or it could be a collection
of them, or a defect, or a lattice phonoon.
The background consists of all remaining degrees of freedom, including the
lattice atoms and any other electrons not treated explicitly.
This distinction is essential for identifying the relevant symmetries and
conserved quantities.

In a crystalline environment, the electronic Hamiltonian is invariant under
\emph{discrete} translations of the electron by lattice vectors, with the
background held fixed.
The associated conserved quantity is \emph{pseudomomentum}
(crystal momentum); for an electron in a periodic potential this is the familiar
Bloch momentum $\hbar\mathbf{k}$.

\begin{figure}
    \centering
    \includegraphics[width=0.85\linewidth]{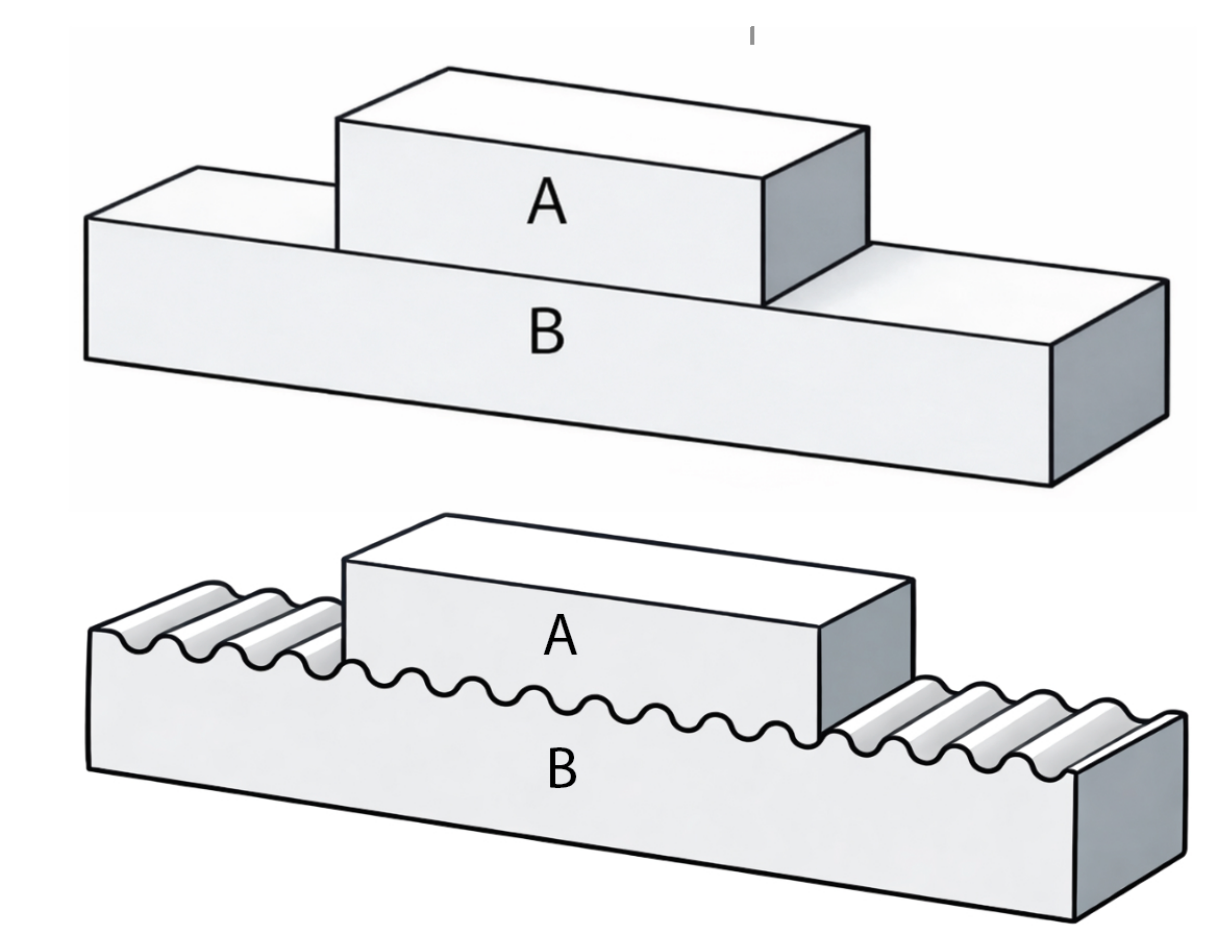}
    \caption{{\bf Sliding Blocks.} Untethered, two blocks attract one another and slide frictionlessly along their common long axis. At the top, both blocks carry conserved mechanical momenta, and clearly $P_{\rm tot}=p_A+p_B$. At the bottom, the same relation holds, but $p_A$ and $p_B$ are now not independently conserved; if the boxes are sliding over one another,their sum nevertheless equals the conserved mechanical momentum. The two blocks serve as surrogates for a foreground electron (A) in a perfect background crystal (B), illustrating that both the foreground and the background acquire pseudomomentum in each other’s presence.}
    \label{fig:forbac}
\end{figure}

The same discrete translational symmetry applies to the background. Translating the entire lattice while holding the {\it electron} fixed leaves the Hamiltonian invariant, and therefore assigns a pseudomomentum to the background as well.
(see Fig.~\ref{fig:forbac}).
If the electron (foreground A) coordinate is held fixed and the entire background
B is translated by a lattice vector, the Hamiltonian is again unchanged.
A consistent application of translational symmetry therefore assigns
pseudomomentum not only to the electron but also to the background, even though
only their sum corresponds to a mechanical momentum.

One might object that the problem can always be formulated in the rest frame of
the background, with only the electron moving.
However, this choice of frame obscures the conservation of total momentum of the
combined system and suppresses the dynamical role of background recoil, which is
essential for understanding local scattering events in the interior of a
perfect crystal.
Let $\hat U(\mathbf h)$ translate \emph{all} coordinates, foreground and
background, by an arbitrary vector $\mathbf h\in\mathbb R^3$.  For an isolated
electron+background system,
\begin{equation}
[\hat U(\mathbf h),H]=0\qquad\text{for all }\mathbf h\in\mathbb R^3,
\end{equation}
so the corresponding symmetry is the conserved \emph{total mechanical} momentum, denoted by  $\hat{\mathbf Q}$.

 In the foreground/background formulation  $\hat{\mathbf Q}$ is obtained
as the sum of two subsystem pseudomomenta. Because $\hat U(\mathbf h)$ factorizes into a translation of the foreground
coordinate and an equal  translation of the background configuration,
\begin{equation}
\hat U(\mathbf h)=\hat U_{\mathrm{f}}(\mathbf h)\,\hat U_{\mathrm{b}}(\mathbf h),
\end{equation}
we have
\begin{equation}
\hat{\mathbf{Q}}
=
\hat{\mathbf P}
+\hat{\mathbf p}
\label{eq:Ktot}
\end{equation}
 Here $\hat{\mathbf p}$ is the pseudomomentum carried by the
foreground electron (the generator of its translation), and
$\hat{\mathbf P}$ is the pseudomomentum carried by the background. The total background pseudomomentum
decomposes into
\begin{equation}
\hat{\mathbf P}
=
\hat{\mathbf P}_{\bf R }
+
\hat{\mathbf P}_{\rm ph}
\end{equation}
 the center-of-mass (zero-mode) contribution
and the internal phonon contribution.
(the generator of translating the background as a whole). Their sum
$\hat{\mathbf Q} =\hat{\mathbf P}+\hat{\mathbf p}$ is the conserved \emph{mechanical} momentum of
the isolated system. For convenience, we can set it to 0; it never changes and we can now ignore it.

Even under the usual rule stating that defects, such as a crystal edge  or impurity, scatter electrons elastically, there is a subtlety within the prior framework: the electron changes its pseudomomentum, but  as is conventionally understood, the lattice cannot  respond or recoil in kind. 
A pseudomomentum cannot be combined with a mechanical
momentum. We have now seen, however, that both foreground and background have pseudomomentum, and can exchange it freely. Total momentum is conserved  for example in the very real deflection of an electron at an edge by equal and opposite pseudomomentum changes. 

Historically, there has been a trend to treat electron scattering in a crystal as if it were Compton scattering in free space. There is no ether, no center of mass of a substrate. This is elegant, but also it misses something important.   Here, we restore for a periodic lattice a full set of rules that apply in free space, as one would wish. There is a total psuedomomentum conserved even if an electron encounters a defect. 
However, unlike in free space, in crystals there is an aether, so to speak, a silent partner: the lattice. It is more than its phonons, it has a center of mass.

\begin{figure}
    \centering
    \includegraphics[width=.51\linewidth]{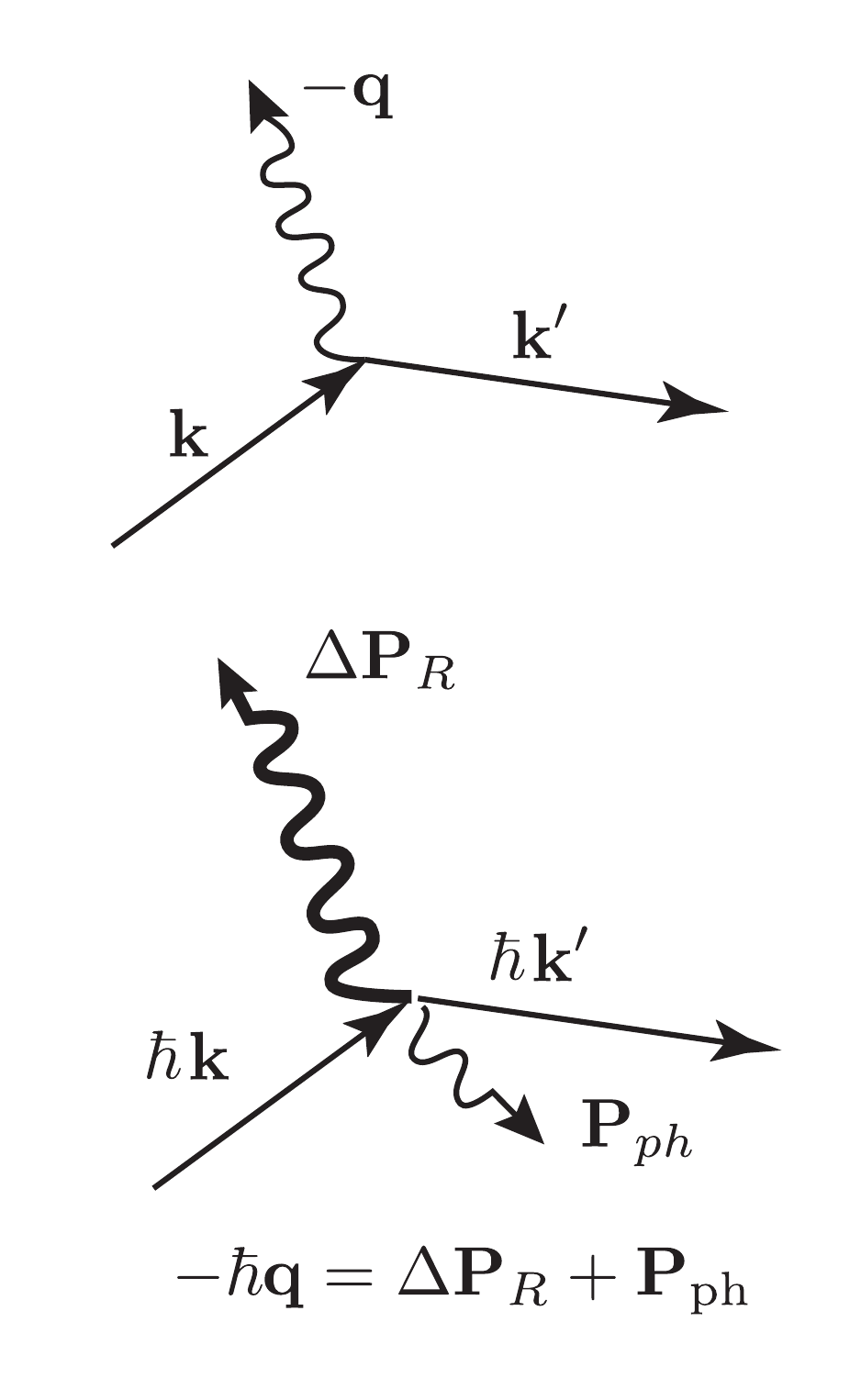}
    \caption{(top) Traditional diagram of electron deflection by phonon emission or absorption. 
(bottom) Momentum transfer in a perfect crystal when the lattice zero mode is retained. 
The collision occurs on a dynamical lattice: quasielastic phonons may be produced, 
while the change in electron momentum is balanced jointly by the phonon momentum and the 
center-of-mass recoil of the lattice.
Note: When momentum $\Delta p=\hbar q$ is transferred to the lattice center of mass, the 
associated de Broglie wavelength is $\lambda=2\pi/q$, identical to that of the scattering 
process. The large lattice mass suppresses the recoil energy but not the wavelength or the 
momentum transfer.}
    
\label{fig:twodagrams}
\end{figure}

Figure \ref{fig:twodagrams} reflects the revised understanding and diagram including the participation of the lattice. This collision takes place in an ``ether.''

\begin{figure}
    \centering
    \includegraphics[width=1.05\linewidth]{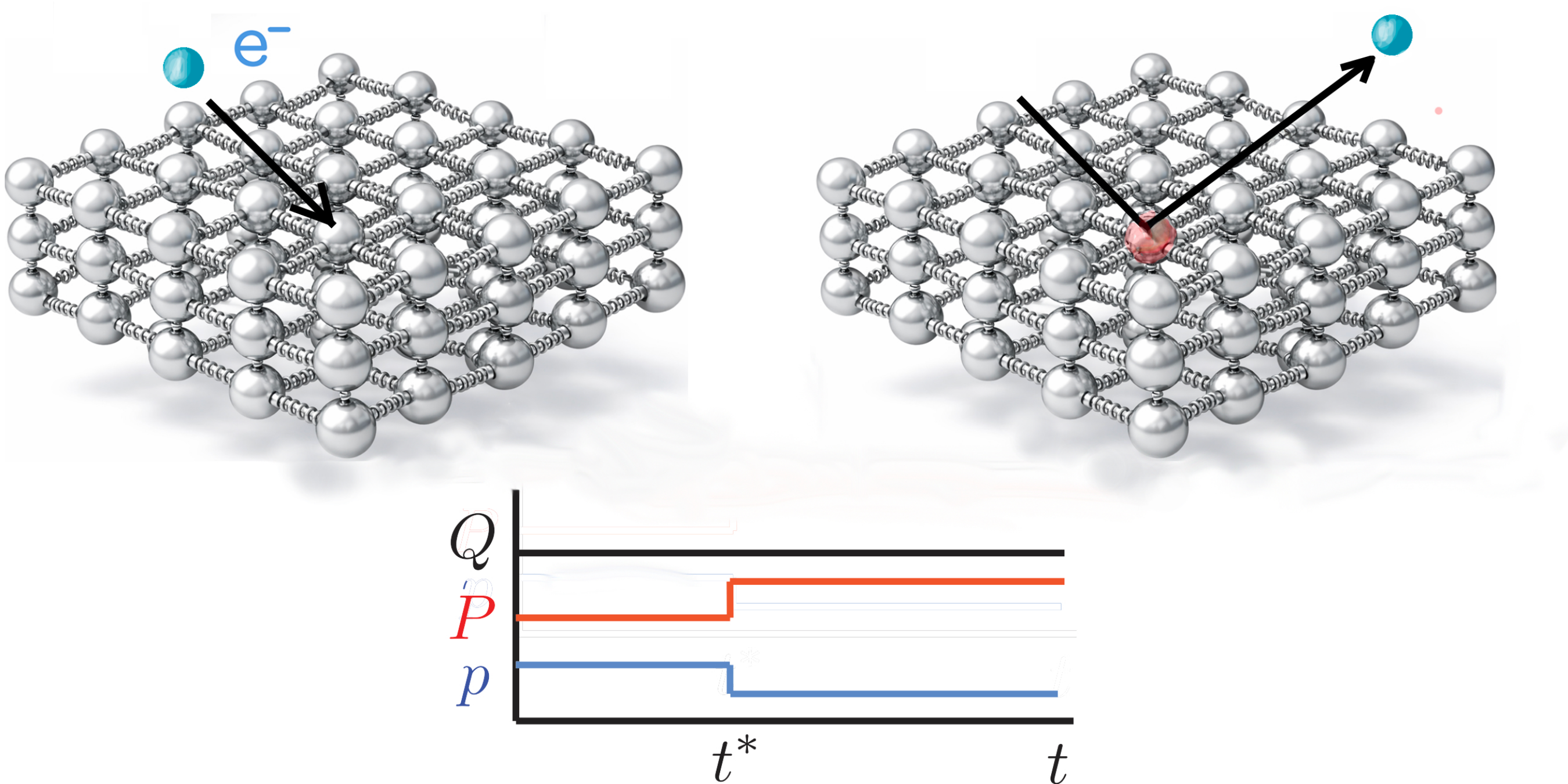}
    \caption{The total momentum (black) remains constant as the foreground (electron, blue) and background (lattice, red) momenta jump abruptly at the collision time $t^*$. There is no delay in the total lattice momentum change, even though only a single atom initially carries the momentum. While this is elementary, we emphasize it because there is a common tendency to associate momentum flow directly with energy flow in the medium. Indeed there is momentum flow among particles, but the total is fixed immediately. The total appears in the wavefunctions. As with M\"ossbauer scattering, the red atom has gained momentum and with it the whole lattice, but when measured, the lattice  may still be in the same vibrational state as before the collision (elastic scattering).}
    \label{fig:ecol}
\end{figure}

The total momentum $\hat{\mathbf Q}$ is conserved, but the foreground and
background momenta, $\hat{\mathbf p}$ and $\hat{\mathbf P}=\hat{\mathbf P}_R + \hat{\mathbf P}_{\rm int}$, undergo equal and
opposite changes, in exact analogy with collisions in free space.
In particular, any change in $\hat{\mathbf p}$ necessarily induces a
compensating change in the global (zero-mode) contribution
$\hat{\mathbf P}_R$, while changes in the internal degrees of freedom are
optional
This exchange does not require the creation or annihilation of internal lattice
excitations and occurs immediately at the collision event
(see Fig.~\ref{fig:ecol}).
The very real recoil of the atom labeled in red in Fig.~\ref{fig:ecol} may very well prove to have excited no vibrational quanta when measured in a number state basis. This is just the elementary fact that a slightly displaced ground state of a harmonic oscillator is still mostly in the ground state, if measured.

In what follows, we formalize a symmetric foreground-background structure,
derive the elastic scattering amplitude within first-order perturbation theory,
and compute the elastic fraction explicitly.
We show that this fraction is close to unity for copper at room temperature.
We also clarify how the present formulation extends the Fr\"ohlich framework
without contradicting its successful predictions, and discuss the implications
for the interpretation of transport phenomena in crystalline solids.


\
.
\subsection{Finite-window symmetry}

In any physically realizable crystal the interaction is periodic only over a
finite spatial extent $L$.
Strict translational invariance is therefore absent at the boundaries, and
crystal momentum is not an exact conserved quantity for the full finite system.
This situation is directly analogous to the textbook particle in a box:
the exact eigenstates are standing waves that do not carry definite momentum,
yet traveling wave packets with well-defined average momentum can be
constructed as superpositions of these eigenstates.
Over times short compared with the traversal time to the boundary,
such wave packets propagate as if momentum were conserved.

The same reasoning applies to a large but finite crystal.
Although global translational symmetry is broken by the boundaries,
the Hamiltonian is locally periodic in the bulk.
Wave packets localized far from the boundaries evolve, for long but finite
times, according to the translationally invariant bulk Hamiltonian.
In this regime one may assign a well-defined crystal pseudomomentum to the
packet, up to corrections that vanish as $L\to\infty$.

This type of approximate symmetry is standard in discussions of the
thermodynamic limit, bulk transport theory, and wave-packet dynamics in
finite systems.
For clarity, we refer to it here as \emph{finite-window symmetry}:
translational invariance holds within a spatial and temporal window set by
the system size.
No new symmetry is implied; rather, this terminology emphasizes that the
use of bulk momentum quantum numbers in a finite crystal is an
approximation controlled by the ratio of the observation time and length
scales to the system size.

In a macroscopic crystal this approximation is exceedingly accurate.
Localized wave packets (“corpuscles”) can be constructed that propagate
nearly as Bloch waves with well-defined pseudomomentum and experience long
travel times before interacting with boundaries.
This construction permits the use of the notation and results associated
with Born–von~K\'arm\'an boundary conditions as a calculational device,
without assuming literal periodicity of the finite sample.

In a finite crystal, the exact normal modes are standing waves. As exact eigenmodes they have zero net mechanical momentum. Any “phonon carrying momentum $\hbar q$” is then not an eigenstate; it’s a wavepacket (a superposition of standing modes) that only behaves like a traveling object transiently until it reflects, mode-mixes, etc. 


\section{Lattice degrees of freedom and zero modes}

We have already encountered a zero mode in the minimal model of an electron and an atom chain. section~\ref{ring}. The mode there is the angular momentum. 

Now we venture out more generally, and find the same large   phase space becomes available for elastic and quasielastic scattering processes, a phase space that is simply absent if the center of mass is absent in the Hamiltonian.

We consider a finite crystal consisting of atoms with coordinates
$\hat{\mathbf R}_j$ and masses $m_j$, with a total lattice mass
$M=\sum_j m_j$, minus the foreground electron.  We define the center-of-mass coordinate of the lattice
background as
\begin{equation}
\hat{\mathbf R}
\equiv
\frac{1}{M}\sum_j m_j\,\hat{\mathbf R}_j .
\end{equation}
It is convenient to decompose the atomic coordinates as
\begin{equation}
\begin{aligned}
\hat{\mathbf R}_j
&=
\hat{\mathbf R}
+
\boldsymbol{\xi}^{(0)}_j
+
\hat{\mathbf u}_j,
\\
\sum_j m_j \boldsymbol{\xi}^{(0)}_j
&=
\mathbf 0,
\qquad
\sum_j m_j \hat{\mathbf u}_j
=
\mathbf 0,
\end{aligned}
\label{eq:decompose_rj}
\end{equation}
where \(\hat{\mathbf R}\) is the exact lattice center-of-mass coordinate,
\(\boldsymbol{\xi}^{(0)}_j\) denote fixed reference positions measured relative
to the center of mass, and \(\hat{\mathbf u}_j\) represent internal,
dynamical lattice displacements.
The mass-weighted constraints ensure that all internal degrees of freedom carry
zero net momentum, so that \(\hat{\mathbf R}\) describes the unique translational
zero mode.

The classical and quantum theory of lattice vibrations explicitly separates
rigid translation from internal vibrations.  Born and Huang showed that the equations
of motion of a finite crystal admit three zero-frequency normal modes
corresponding to rigid translation of the crystal as a whole, while the internal
normal modes satisfy the mass-weighted constraint $
    \sum_j m_j \hat{\mathbf u}_j = \mathbf 0,$
so that internal vibrations carry no net mechanical momentum
\cite{BornHuang1954}.  All mechanical   momentum of the crystal resides in
the translational zero mode.



%
\section{Electron-Lattice interaction}
\subsection{Foreground and background Hamiltonians}

.
We write the total Hamiltonian as
\begin{equation}
H = H_{\rm fg} + H_{\rm bg} + H_{\rm int},
\end{equation}
where \(H_{\rm fg}\) describes the foreground electron,
\(H_{\rm bg}\) the background degrees of freedom,
and \(H_{\rm int}\) their coupling.

For a single foreground electron we take
\begin{equation}
H_{\rm fg} = \frac{\hat{\mathbf p}^2}{2m},
\end{equation}
where \(\hat{\mathbf p}\) denotes the generator of translations of the
foreground coordinate \(\hat{\mathbf r}\).
The use of a bare kinetic-energy operator reflects the fact that the electron
has not been pre-diagonalized into Bloch states.
All effects of the periodic crystal potential, including band structure,
pseudomomentum exchange, and scattering, are  generated 
through the interaction Hamiltonian \(H_{\rm int}\).
This is the same starting point adopted by Fr\"ohlich~\cite{Frohlich1954}.

The remaining electrons of the metal are treated as part of the background.
Together with the ions they establish the effective lattice potential, provide
screening, and ensure mechanical stability.
Their degrees of freedom are therefore implicit in \(H_{\rm bg}\).

\subsection{Density operator and exact factorization}

The lattice density operator admits an exact factorization into a
center-of-mass translation and an internal part:
\begin{equation}
\begin{aligned}
\hat\rho_{\mathbf q}
&=
\sum_j e^{-i\mathbf q\cdot \hat{\mathbf R}_j}
=
e^{-i\mathbf q\cdot \hat{\mathbf R}}\,
\hat\rho^{\rm int}_{\mathbf q},
\\
\hat\rho^{\rm int}_{\mathbf q}
&=
\sum_j
e^{-i\mathbf q\cdot(\boldsymbol{\xi}^{(0)}_j+\hat{\mathbf u}_j)} .
\end{aligned}
\label{eq:rho_factorization}
\end{equation}
Here $\hat{\mathbf R}$ is the lattice center-of-mass operator,
$\boldsymbol{\xi}^{(0)}_j$ are equilibrium positions,
and $\hat{\mathbf u}_j$ are internal displacements.
Equation~\eqref{eq:rho_factorization} is an exact operator identity,
independent of any choice of basis (plane waves, Bloch waves,
or standing waves).

This identity expresses a purely kinematic fact:
a rigid translation of the lattice multiplies every density
Fourier component by the phase
$e^{-i\mathbf q\cdot \hat{\mathbf R}}$,
while all internal lattice dynamics are contained entirely in
$\hat\rho^{\rm int}_{\mathbf q}$.

\subsection{Density–density interaction and pseudomomentum conservation}

The electron–lattice interaction may be written in density–density form as
\begin{equation}
H_{\rm int}
=
\sum_{\mathbf q} V(\mathbf q)\,
\rho_{\rm fg}(\mathbf q)\,
\rho_{\rm bg}(-\mathbf q),
\label{eq:Hdd_unified}
\end{equation}
following the formalism of Van Hove~\cite{VanHove1954}.
No assumption about Bloch-wave eigenstates is required;
$\mathbf q$ labels Fourier components of the interaction kernel. This equation is the essence of the term: density-density interaction. 

For a tagged foreground electron, we write the aperiodic part of the Bloch waves as 
\begin{equation}
\rho_{\rm fg}(\mathbf q)
=
e^{i\mathbf q\cdot\hat{\mathbf r}}.
\end{equation}

Substituting into Eq.~\eqref{eq:Hdd_unified} gives
\begin{equation}
\boxed{
H_{\rm int}
=
\sum_{\mathbf q}
V(\mathbf q)\,
e^{i\mathbf q\cdot(\hat{\mathbf r}-\hat{\mathbf R})}\,
\hat\rho^{\rm int}_{-\mathbf q}
}
\label{eq:Hint_relcoord_unified}
\end{equation}
Equation~\eqref{eq:Hint_relcoord_unified} shows that the interaction depends
only on the relative coordinate
\(
\hat{\bm\rho}=\hat{\mathbf r}-\hat{\mathbf R}.
\)
The symmetry of the phase factor
\(e^{i\mathbf q\cdot(\hat{\mathbf r}-\hat{\mathbf R})}\)
is the central structural feature:
pseudomomentum $\mathbf q$ is exchanged between foreground and background
with change`s always equal and opposite.

    \vskip .1in
When the electron momentum changes from $k$ to $k'$, the lattice
background absorbs the compensating momentum
\begin{equation}
\Delta P_{\rm bg} = -\hbar (k'-k).
\end{equation}
This momentum is carried entirely by the center-of-mass (zero mode) of
the background. Internal lattice excitations describe only relative
motion and carry no net mechanical momentum.

The associated energy change separates into internal and recoil
contributions,
\begin{equation}
\Delta E_{\rm lat}
=
\Delta E_{\rm int}
+
\frac{\Delta P_R^{\,2}}{2M_{\rm tot}},
\end{equation}
where $\Delta P_R = -\hbar(k'-k)$ is the center-of-mass recoil momentum.
For macroscopic $M_{\rm tot}$, the recoil energy is negligible, so large
momentum transfer can occur with little or no change in internal lattice
energy.


Translational invariance implies that the background degrees of freedom
contain a rigid (center-of-mass) coordinate $\mathbf R$ with conjugate
momentum
\[
\hat{\mathbf P}_{\bf R} = -\,i\hbar \frac{\partial}{\partial \mathbf R}.
\]
The remaining background variables describe internal excitations
(phonons or distortions) defined relative to the center of mass.
Accordingly, the background Hilbert space factorizes as
\[
\mathcal H_{\rm bg}
=
\mathcal H_R \otimes \mathcal H_{\rm int}.
\]
All mechanical momentum of the background resides in the zero mode,
\begin{equation}
\hat{\mathbf P} = \hat{\mathbf P}_{\bf R},
\end{equation}
while the internal degrees of freedom carry no net mechanical momentum.

The interaction operator
\[
e^{i\mathbf q\cdot(\hat{\mathbf r}-\hat{\mathbf R})}
\]
transfers momentum $\hbar\mathbf q$ from the electron to the background
as a whole. This momentum is carried entirely by the center-of-mass
degree of freedom. Lattice recoil is therefore not an additional
dynamical channel but a direct consequence of translational symmetry.

A nonvanishing matrix element of $H_{\rm int}$ requires
\begin{equation}
\Delta\mathbf p = +\hbar\mathbf q,
\qquad
\Delta\mathbf P = -\hbar\mathbf q,
\end{equation}
so that the total mechanical momentum
\(
\hat{\mathbf Q}=\hat{\mathbf p}+\hat{\mathbf P}
\)
is conserved exactly.

Crucially, this momentum exchange is independent of phonon occupation.
Elastic matrix elements of $\hat\rho^{\rm int}_{-\mathbf q}$ are
generically nonzero, so the lattice can absorb finite momentum through
center-of-mass recoil without any change in its internal state.
\section{Elastic internal scattering and the Debye-Waller fraction}

We now show that, even within a strictly perturbative treatment
based on Bloch electrons and phonon number states,
robust elastic scattering channels exist already at first order in the
electron–lattice coupling.
These channels become explicit once the full lattice density operator is
retained and the translational zero mode is treated as a dynamical degree of freedom.
The resulting processes are elastic with respect to internal lattice
excitations, despite involving momentum exchange with the lattice as a whole.

\subsection{Emergence of the Bloch potential}

Separating the background density operator into its equilibrium
expectation value (taken in the lattice center-of-mass frame)
and fluctuations,
\begin{equation}
\rho_{\rm bg}(-\mathbf q)
=
\langle\rho_{\rm bg}(-\mathbf q)\rangle
+
\delta\rho_{\rm bg}(-\mathbf q),
\end{equation}
the static component produces an effective periodic potential
\begin{equation}
U_{\rm Bloch}(\hat{\mathbf r})
=
\sum_{\mathbf G}
V(\mathbf G)\,
e^{i\mathbf G\cdot\hat{\mathbf r}}\,
\langle\rho_{\rm bg}(-\mathbf G)\rangle,
\label{eq:U_Bloch_unified}
\end{equation}
where $\mathbf G$ are reciprocal lattice vectors.

The corresponding foreground Hamiltonian becomes
\begin{equation}
H_{\rm fg}
=
\frac{\hat{\mathbf p}^2}{2m}
+
U_{\rm Bloch}(\hat{\mathbf r}),
\end{equation}
which is the usual Bloch Hamiltonian. In this way, the periodic
potential is not imposed \emph{a priori}, but emerges as the
mean-field component of the translationally invariant
density-density interaction in the background center-of-mass frame.

The remaining part of $H_{\rm int}$ describes coupling to fluctuations
of the background density. Crucially, it retains the explicit
center-of-mass translation factor, ensuring that momentum transfer is
accounted for by recoil of the background as a whole, rather than being
assigned solely to internal phonon modes.
\subsection{Initial and final states}

For a perturbative treatment, we consider product states
\begin{equation}
\Psi_i
=
\Psi_{n\mathbf k}(\mathbf r)\,
\Psi^{\rm bg}_i,
\qquad
\Psi_f
=
\Psi_{n'\mathbf k'}(\mathbf r)\,
\Psi^{\rm bg}_f,
\end{equation}
where $\Psi_{n\mathbf k}$ are extended-zone Bloch eigenstates of the
static lattice Hamiltonian and $\Psi^{\rm bg}_{i,f}$ are arbitrary
background states in the full lattice Hilbert space (not restricted to
phonon number eigenstates).

\subsection{Interaction and matrix element}

From Eq.~\eqref{eq:Hint_relcoord_unified}, the interaction is
\begin{equation}
H_{\rm int}
=
\sum_{\mathbf q}
V(\mathbf q)\,
e^{i\mathbf q\cdot(\hat{\mathbf r}-\hat{\mathbf R})}
\hat\rho^{\rm int}_{-\mathbf q},
\end{equation}
where $\hat{\mathbf R}$ is the lattice center-of-mass operator and
$\hat\rho^{\rm int}_{-\mathbf q}$ acts on internal background degrees of
freedom.

Between product states the interaction matrix element is
\begin{align}
\mathcal M_{fi}
&=
\big\langle n'\mathbf k',\Psi_f^{\rm bg}\big|
H_{\rm int}
\big|n\mathbf k,\Psi_i^{\rm bg}\big\rangle \nonumber
\\
&=
V(\mathbf q)\,
\langle n'\mathbf k'|
e^{i\mathbf q\cdot\hat{\mathbf r}}
|n\mathbf k\rangle
\,
\langle \Psi_f^{\rm bg}|
e^{-i\mathbf q\cdot\hat{\mathbf R}}
\hat\rho^{\rm int}_{-\mathbf q}
|\Psi_i^{\rm bg}\rangle .
\end{align}
\subsection{Total pseudomomentum conservation}

Translational invariance of the full electron-lattice system implies
exact conservation of total pseudomomentum,
\begin{equation}
\mathbf k' - \mathbf k
=
- \big(
\mathbf K^{\rm bg}_f - \mathbf K^{\rm bg}_i
\big),
\label{eq:psm_exact_unified}
\end{equation}
where $\mathbf K^{\rm bg}_{i,f}$ are eigenvalues of the total background
pseudomomentum operator, including the zero (center-of-mass) mode.

Equation~\eqref{eq:psm_exact_unified} is the only kinematic constraint
imposed by translational symmetry. No additional reciprocal lattice
vector is required when electronic states are treated in the
extended-zone scheme.

In mechanical form this reads
\begin{equation}
\Delta\mathbf p
=
-\Delta\mathbf P,
\end{equation}
so that the total mechanical momentum
\(
\hat{\mathbf Q}=\hat{\mathbf p}+\hat{\mathbf P}
\)
is conserved exactly.

\section{Partition of pseudomomentum transfer}

We now examine how the transferred pseudomomentum
\(
\hbar(\mathbf k'-\mathbf k)
\)
is stored within the background.

The background matrix element contains two distinct structures,
\begin{equation}
e^{-i\mathbf q\cdot\hat{\mathbf R}}
\qquad \text{(center-of-mass recoil)},
\end{equation}
and
\begin{equation}
\hat\rho^{\rm int}_{-\mathbf q}
\qquad \text{(internal lattice operator)}.
\end{equation}

The conservation law \eqref{eq:psm_exact_unified} constrains only the
total background pseudomomentum change
\(
\Delta\mathbf K^{\rm bg}.
\)
It places no requirement on how that momentum is distributed between
internal modes and the center-of-mass degree of freedom.

\subsection{Conventional phonon bookkeeping}

In the clamped-lattice approximation one neglects the recoil operator
and expands background states in phonon number eigenstates. In that
restricted description one identifies
\begin{equation}
\Delta\mathbf K^{\rm bg}
=
\sum_{\mathbf q}
\Delta n_{\mathbf q}\,\mathbf q,
\end{equation}
so that electronic momentum change is compensated entirely by phonon
creation or annihilation.

Energy conservation then requires
\begin{equation}
\epsilon_{n'\mathbf k'}
=
\epsilon_{n\mathbf k}
\pm
\hbar\omega_{\mathbf q\lambda}.
\end{equation}

This phonon-counting condition is therefore a consequence of the
harmonic number-state basis together with suppression of the
center-of-mass degree of freedom. It restricts the dynamics to a
low-codimension subset of the full many-body phase space, in which
momentum and energy must be matched by discrete phonon quanta.
\subsection{Full Hilbert-space description}

In the complete electron-lattice Hilbert space,
Eq.~\eqref{eq:psm_exact_unified} remains the only exact constraint.
The compensating background momentum is carried entirely by the
center-of-mass (zero mode) of the lattice. Internal degrees of freedom
describe only relative motion and carry no net mechanical momentum.

Thus, the association of background momentum transfer with specific
phonon occupation changes is not fundamental, but a consequence of
discarding the zero mode. Phonons may accompany a scattering event, but
they do not provide the required momentum balance. The zero mode is the
unique carrier of mechanical momentum.

In the clamped-lattice formulation, an electronic transition
$k \to k'$ is restricted to those phonon modes whose wavevector and
frequency simultaneously satisfy both momentum and energy matching
conditions. In low dimensions or at low temperature, such solutions are
sparse and may be absent altogether except through Umklapp. When it is
said that Umklapp momentum is ``absorbed by the lattice,'' the role of
the zero mode is implicitly reintroduced for the special case of a
reciprocal lattice vector $G$, even though the lattice can in fact absorb
arbitrary momentum.

By contrast, when the translational zero mode is retained, momentum
conservation is enforced by recoil of the background as a whole, while
energy conservation is governed independently by the internal spectrum.
The internal lattice states form an exponentially dense many-body
manifold, and all configurations within an energy window
$|E_f - E_i| \lesssim \Gamma$
contribute. The number of quasi-elastic channels therefore scales with
the many-body density of states and becomes macroscopic in the
thermodynamic limit. This vast phase space is absent when the zero mode
is omitted.
\subsection{Recoil-dominated channel}

When the zero recoil mode is retained, the required pseudomomentum
transfer is taken up entirely by the background center-of-mass degree
of freedom:
\begin{equation}
\langle \mathbf K'|
e^{-i\mathbf q\cdot\hat{\mathbf R}}
|\mathbf K\rangle
=
\delta_{\mathbf K',\,\mathbf K-\mathbf q},
\end{equation}
so that total pseudomomentum conservation
\(
\mathbf k' - \mathbf k
=
- (\mathbf K^{\rm bg}_f - \mathbf K^{\rm bg}_i)
\)
is satisfied without reference to phonon occupation changes.

The associated recoil energy is
\begin{equation}
\Delta E_{\rm recoil}
=
\frac{\hbar^2}{2M_{\rm tot}}
\left[
(\mathbf K-\mathbf q)^2-\mathbf K^2
\right]
=
\frac{(\hbar q)^2}{2M_{\rm tot}}
-\frac{\hbar^2}{M_{\rm tot}}\mathbf K\!\cdot\!\mathbf q,
\end{equation}
which scales as $1/M_{\rm tot}$ and therefore vanishes in the
thermodynamic limit.
To leading order in $1/N$, the scattering is thus elastic with respect
to internal lattice excitations.

The internal factor
\begin{equation}
\langle \Psi_f^{\rm bg}|
\hat\rho^{\rm int}_{-\mathbf q}
|\Psi_i^{\rm bg}\rangle
\end{equation}
remains finite even when no phonon quanta are created or annihilated.
For a thermally fluctuating background its diagonal contribution
reduces, upon averaging, to the Debye-Waller factor
\(
e^{-W(\mathbf q,T)}.
\)

\subsection{Quasielastic and superelastic processes}

More generally,
\(
\hat\rho^{\rm int}_{-\mathbf q}
\)
contains all orders in internal lattice operators.
Transitions therefore exist in which the center-of-mass absorbs the
entire pseudomomentum transfer, while the internal modes undergo small
changes that adjust the energy balance. Such processes are quasielastic:
\begin{equation}
\epsilon_{n'\mathbf k'}
\approx
\epsilon_{n\mathbf k},
\qquad
|\Delta E_{\rm int}|
\ll
\epsilon_{n\mathbf k},
\end{equation}
with the exact kinematic constraint always enforced by
center-of-mass recoil.

At finite temperature, superelastic channels occur when
pre-existing internal excitations are partially annihilated,
transferring energy to the electron while the net pseudomomentum
remains balanced by the center-of-mass degree of freedom.

\bigskip

\subsection{Elastic and quasi-elastic processes beyond the Fr\"ohlich term}

Consider the electron-lattice interaction in its full density-density form
\begin{equation}
H_{\rm int}
=
\sum_q V(q)\,\rho_e(-q)\,\rho_{\rm bg}(q),
\qquad
\rho_e(-q)=\sum_k c^\dagger_{k+q}c_k .
\end{equation}
The background density may be written as
\begin{equation}
\begin{aligned}
\rho_{\rm bg}(q)
&=
\sum_j e^{-iq(R_j+u_j)} \\
&=
\sum_j e^{-iqR_j}
\left(
1-iq\,u_j-\frac{q^2}{2}u_j^2+\cdots
\right).
\end{aligned}
\end{equation}
where $R_j$ are equilibrium positions and $u_j$ displacements.

\begin{figure}
     \centering
\includegraphics[width=0.95\linewidth]{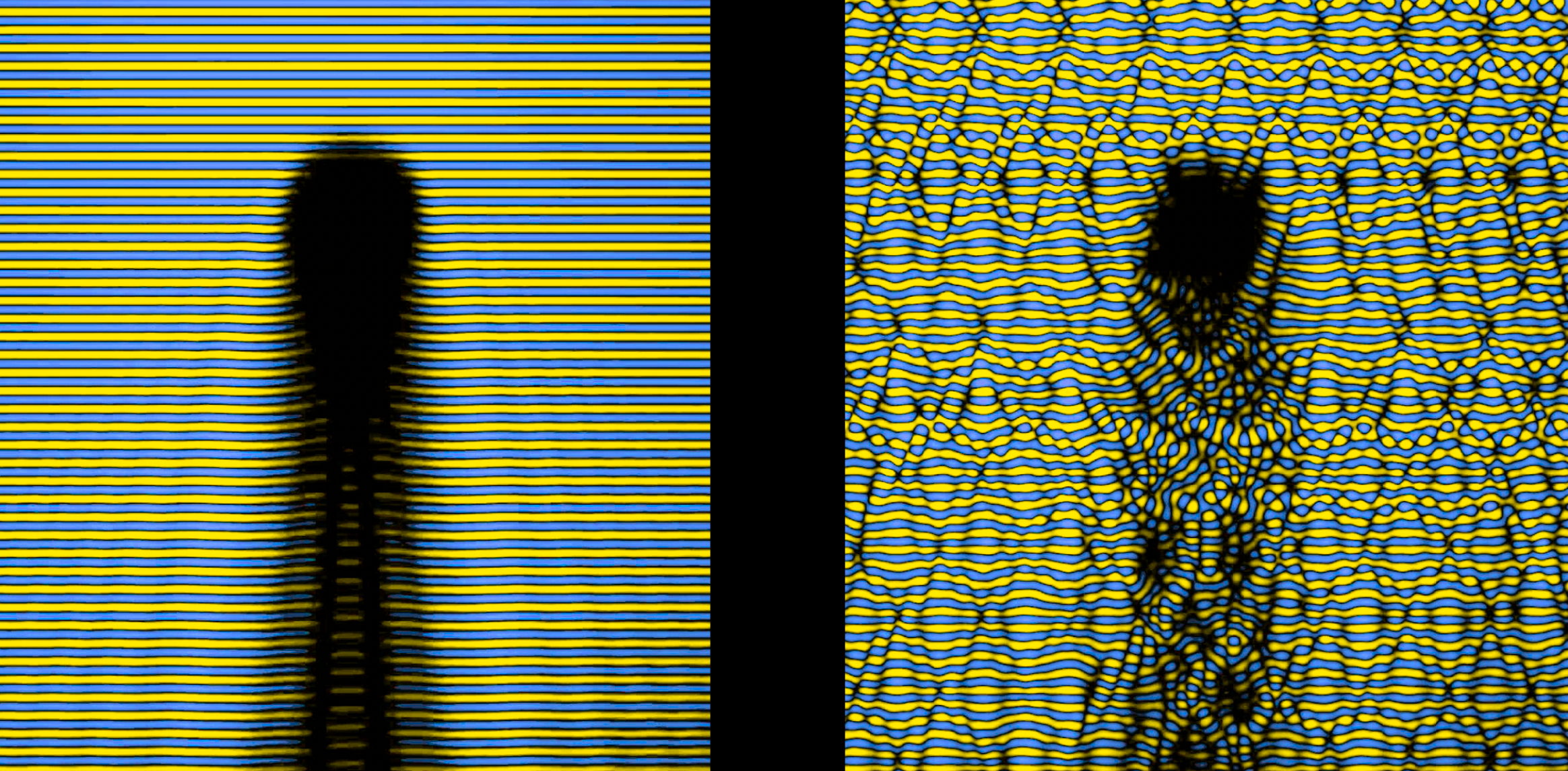}
     \caption{\textbf{Interrupted Bloch wave} Both panels feature a circular  absorbing zone seen clearly, as the real part of a plain wave (left) and of a Bloch wave (right) enter from above.  A shadow persists and slowly fills by diffraction on the left, plane wave case, but   strong scattering by the triangular lattice present on the right quickly fills in the void. This scattering, which is obscured by periodic coherence in the unobstructed Bloch waves, is plainly revealed as the void fills.  The  point is that  the strong unprotected scattering in the shadow region cannot be strongly inelastic, or else passive barriers in periodic lattices would cause  heating downstream. }
    \label{Kohnspoint}
 \end{figure}

The zeroth-order term gives
\begin{equation}
H^{(0)}
=
\sum_{k,q} V(q)\,c^\dagger_{k+q}c_k
\sum_j e^{-iqR_j},
\end{equation}
which contains no phonon operators and is therefore purely elastic. In a
perfect periodic crystal this reduces to reciprocal lattice scattering
($q=G$), i.e.\ Bragg processes.
It is responsible for the static band structure. As Kohn noted\cite{Kohn1964}, there is plenty of elastic deflection going into the construction of bands, which gets masked in making Bloch waves.  Figure~\ref{Kohnspoint} makes this clear. 
 
The linear term
\begin{equation}
H^{(1)}
=
-i\sum_{k,q} q\,V(q)\,c^\dagger_{k+q}c_k
\sum_j e^{-iqR_j}u_j,
\end{equation}
yields, upon expanding $u_j$ in phonon operators, the familiar
Fr\"ohlich interaction in which each term carries a definite phonon
pseudomomentum.

At quadratic order one finds
\begin{equation}
H^{(2)}
=
-\frac12\sum_{k,q} q^2 V(q)\,c^\dagger_{k+q}c_k
\sum_j e^{-iqR_j}u_j^2,
\end{equation}
and writing
\begin{equation}
u_j=\frac{1}{\sqrt N}\sum_Q U_Q\,e^{iQR_j}(a_Q+a^\dagger_{-Q}),
\end{equation}
one obtains terms of the form
\begin{equation}
\begin{split}
H^{(2)}
\sim
\sum_{k,q}\sum_{Q,Q'}
c^\dagger_{k+q}c_k\,
(a_Q+a^\dagger_{-Q})(a_{Q'}+a^\dagger_{-Q'}) \\
\times
\sum_j e^{-i(q-Q-Q')R_j}.
\end{split}
\end{equation}
Among these are phonon-diagonal contributions with $Q'=-Q$,
\begin{equation}
H^{(2)}_{\rm el}
\sim
\sum_{k,q,Q}
c^\dagger_{k+q}c_k\,
\bigl(2a_Q^\dagger a_Q+1\bigr)
\sum_j e^{-iqR_j},
\end{equation}
which have nonzero matrix elements
\begin{equation}
\langle \nu|H^{(2)}_{\rm el}|\nu\rangle \neq 0.
\end{equation}
These describe elastic electron scattering with $\nu\to\nu$, and more
generally quasi-elastic processes when $\nu\to\nu'$ with small energy
change.

Thus, once the interaction is treated beyond the linear Fr\"ohlich
approximation, elastic and quasi-elastic channels appear already {\it at first
order} in the density-density interaction. In such processes
$\Delta K_{\rm ph}=0$, while in general $\Delta K_e\neq Gm$, so the
standard selection rule
\begin{equation}
\Delta K_e+\Delta K_{\rm ph}=Gm
\end{equation}
is not satisfied within the zero-mode-removed theory.

This is consistent with the observation, emphasized by Kohn, that the
electronic band structure itself arises from coherent elastic scattering
from the periodic lattice potential. Elastic scattering is therefore not
an exceptional process but the dominant one that defines the electronic
states. The restriction to inelastic, one-phonon processes is a feature
of the linearized phonon representation, not of the underlying dynamics.

The appearance of elastic and quasi-elastic channels thus exposes the
limitations of the reduced description. Once the lattice center-of-mass
degree of freedom is omitted, neither mechanical momentum nor
pseudomomentum is properly accounted for within the phonon subspace
alone. The full problem requires a dynamical background capable of
carrying the compensating momentum.


\subsection{Thermal averaging and the Debye-Waller fraction}

In thermal equilibrium the internal lattice degrees of freedom are described by
the density matrix
\begin{equation}
\hat{\varrho}
=
\frac{e^{-\beta \hat H_{\mathrm{lat}}}}{Z}
=
\sum_{n} p_n\,|n\rangle\langle n|,
\qquad
p_n=\frac{e^{-\beta E_n}}{Z},
\end{equation}
where $|n\rangle$ denote exact many-body eigenstates of the lattice Hamiltonian
$\hat H_{\mathrm{lat}}$ with energies $E_n$. For elastic internal scattering the relevant quantity is the
thermal expectation value of the internal density operator,
\begin{equation}
\langle \rho^{\rm int}_{\mathbf q}\rangle_T
=
\sum_n p_n\,
\langle n|\rho^{\rm int}_{\mathbf q}|n\rangle .
\end{equation}

The coherent elastic scattering intensity is
\begin{equation}
I_{\mathrm{el}}(\mathbf q)
=
\big|\langle \rho^{\rm int}_{\mathbf q}\rangle_T\big|^2 .
\end{equation}

The total scattering intensity summed over all final internal states is
\begin{equation}
I(\mathbf q)
=
\left\langle
\rho^{\rm int}_{\mathbf q}
\rho^{\rm int}_{-\mathbf q}
\right\rangle_T .
\end{equation}

The elastic (coherent) branching fraction is therefore
\begin{equation}
f_{\mathrm{el}}(\mathbf q,T)
=
\frac{
\big|\langle \rho^{\rm int}_{\mathbf q}\rangle_T\big|^2
}{
\left\langle
\rho^{\rm int}_{\mathbf q}
\rho^{\rm int}_{-\mathbf q}
\right\rangle_T
}.
\end{equation}

For a harmonic lattice one finds\cite{LandauStatMech,LoveseyNeutronScattering}
\begin{equation}
f_{\mathrm{el}}(\mathbf q,T)
\simeq
e^{-2W(\mathbf q,T)},
\qquad
W(\mathbf q,T)
=
\frac{1}{2}
\left\langle
(\mathbf q\cdot \hat{\mathbf u})^2
\right\rangle_T ,
\end{equation}
the familiar Debye-Waller result.
Even at finite temperature a substantial fraction of momentum-transfer
events leave the lattice in the same internal state.


\subsection{Estimate of the elastic internal fraction at representative momentum transfers}

For momentum relaxation in a metal it is natural to consider momentum transfers
of order the Fermi momentum, since large-angle scattering dominates the transport
relaxation rate through the usual \((1-\cos\theta)\) weighting.
For copper, a free-electron estimate yields
\begin{equation}
k_F \simeq 1.36\ \text{\AA}^{-1},
\qquad
2k_F \simeq 2.7\ \text{\AA}^{-1}.
\end{equation}
We therefore evaluate the elastic internal (phonon-diagonal) fraction
\(f_{\mathrm{el}}(\mathbf q,T)\) at representative values
\(q \ll k_F\), \(q\sim k_F\), and \(q\sim 2k_F\).

Within the isotropic Debye-Waller estimate used above,
\begin{equation}
2W(\mathbf q,T)
=
\frac{q^2\langle u^2\rangle_T}{3},
\qquad
f_{\mathrm{el}}(\mathbf q,T)\simeq e^{-2W(\mathbf q,T)} .
\end{equation}
Using the room-temperature mean-square displacement for copper,
\(\langle u^2\rangle_{300\mathrm{K}}\simeq 1.9\times10^{-2}\ \text{\AA}^2\),
one finds the values summarized in Table~\ref{tab:fel_cu}.

\begin{table}
\centering
\begin{tabular}{c c c}
\hline
$q$ (\AA$^{-1}$) & $2W$ & $f_{\mathrm{el}}$ \\
\hline
$0.3$ & $5.7\times10^{-4}$ & $0.9994$ \\
$k_F \simeq 1.36$ & $1.2\times10^{-2}$ & $0.988$ \\
$2k_F \simeq 2.7$ & $4.6\times10^{-2}$ & $0.955$ \\
\hline
\end{tabular}
\caption{Estimated elastic internal fraction \(f_{\mathrm{el}}\) for copper at
room temperature (\(T=300\) K) for representative momentum transfers.
Even for large-angle scattering with \(q\sim 2k_F\), the majority of
momentum-relaxing events remain phonon-diagonal, with momentum absorbed
predominantly by the lattice background rather than by internal phonon
excitation.}
\label{tab:fel_cu}
\end{table}

These estimates show that, for momentum transfers relevant to transport,
the overwhelming fraction of scattering events are internally elastic,
in the sense that they leave the lattice in the same vibrational state.
Phonon creation constitutes only a minor branching channel at room temperature,
even for large-angle scattering.
Momentum exchange with the lattice therefore occurs predominantly through
elastic or quasi-elastic coupling to the background degrees of freedom,
with internal phonon excitation representing a secondary correction.

We can motivate this result as follows: For a Cu atom modeled as an Einstein oscillator, a sudden momentum transfer
$\Delta p=\hbar q$ leaves the oscillator in its ground state with probability
\begin{equation}
P_{0\to 0}=\left|\langle 0|e^{iqx}|0\rangle\right|^2
=\exp\!\left(-\frac{E_R}{\hbar\omega_E}\right),
\qquad
E_R=\frac{(\hbar q)^2}{2M}.
\end{equation}
Taking $q=2k_F$ for copper, with $k_F\approx 1.36\,\mathrm{\AA}^{-1}$ and
$M\approx 63.5\,u$, gives $E_R\approx 0.24\,\mathrm{meV}$. For a typical
local vibrational energy $\hbar\omega_E\sim 20$-$30\,\mathrm{meV}$, one finds
$P_{0\to 0}\approx 0.99$.

\section{Experiments suggesting large elastic contributions}

We have already touched on pure wire resistivity and persistent currents above,  as  allowing a more satisfactory explanation of the experiments through inclusion of the zero modes and the expansion to elastic and quasielastic channels that they enable.

Other experimental evidence suggesting predominantly elastic momentum relaxation in metals
at low temperature has accumulated for decades, particularly in the
mesoscopic-transport and quantum-interference literature.
In this section we show that the framework developed here can resolve several
longstanding disparities between theory and experiment regarding electronic transport.
We review a range of experiments that demonstrate a clear separation between
momentum relaxation and energy relaxation, consistent with elastic or quasi-elastic scattering dominating momentum randomization.

\subsection{Hierarchy of momentum and energy relaxation times}

Electrical resistivity and optical conductivity probe momentum relaxation,
quantum-interference phenomena such as weak localization probe phase coherence,
and nonequilibrium measurements including shot noise, Johnson-noise thermometry,
and hot-electron relaxation directly access energy exchange.
Strikingly, these probes consistently reveal a strong hierarchy of timescales,
\begin{equation}
\tau_{\mathrm p} \ll \tau_E ,
\end{equation}
established using independent experimental techniques and across a wide range
of materials.
Here $\tau_{\mathrm p}$ is the timescale on which electronic momentum (or
pseudomomentum) is randomized by scattering, while $\tau_E$ is the timescale
over which electrons exchange energy with the background and relax toward
thermal equilibrium.

The consistent appearance of this hierarchy demonstrates that momentum
relaxation and energy relaxation are distinct physical processes.
These experiments show that electronic momentum can be
efficiently scrambled by a time-dependent environment without significant
energy exchange.
In other words, the lattice need not act as an energy sink in order to act as a momentum sink.
This separation is precisely what is expected when electrons scatter quasi-elastically
from a dynamically fluctuating lattice background.

\subsection{Quantum interference and dephasing}

The suppression of quantum-interference effects with increasing temperature is
commonly described in terms of inelastic scattering processes that exchange
energy with environmental degrees of freedom
\cite{LeeRamakrishnan1985,AltshulerAronovKhmelnitskii1982}.
It is useful, however, to distinguish between \emph{dephasing} as operationally
defined in transport experiments and irreversible \emph{decoherence} in the
strict quantum-mechanical sense.

A broad class of experiments indicates that substantial suppression of
interference can occur even when direct energy relaxation to the lattice
remains weak.
Mesoscopic interference measurements, for example, show a pronounced reduction—
and in some cases an apparent saturation—of the phase-coherence time
$\tau_\phi$ at low temperatures, while independent probes suggest that
electron–phonon energy exchange rates continue to decrease in the same regime
\cite{MohantyJariwalaWebb1997,Imry2002,vonDelft2005}.
This empirical separation of timescales illustrates that phase and momentum
randomization need not scale directly with the rate of substantial energy
transfer to internal lattice excitations.

Weak-localization is robust in experiments, 
implying that the backscattering is coherent and therefore elastic.
\cite{LeeRamakrishnan1985,MohantyJariwalaWebb1997}.Elastic scattering amplitudes, including diffuse non-Bragg amplitudes, are reduced by the well known Debye–Waller factor.

\subsection{Hot-electron relaxation}

A particularly direct separation of momentum and energy relaxation is
provided by hot-electron experiments in metals
\cite{Wellstood1994,Pothier1997,Gershenson2001,Giazotto2006}.
In these measurements, nonequilibrium electron distributions are injected,
and the subsequent momentum randomization and energy loss are probed
independently.
Over broad temperature ranges, the momentum-relaxation time inferred from
transport or magnetotransport is found to be substantially shorter than
the energy-relaxation time associated with thermalization.
Electrons therefore undergo many momentum-deflecting collisions while
remaining close to isoenergetic on the same timescale.

This pronounced separation of timescales indicates that efficient
momentum randomization does not require comparably rapid energy transfer
to lattice excitations.
While inelastic electron–phonon scattering ultimately governs thermalization,
the observed hierarchy suggests that additional scattering channels,
with weak net energy exchange on electronic scales, can play an important
role in momentum relaxation within the metallic state.

Johnson-noise thermometry provides a direct probe of electron–lattice
energy exchange. Experiments consistently show that the electron
temperature relaxes on timescales much longer than those associated with
momentum relaxation inferred from transport\cite{Gershenson2001}. This separation,
$\tau_p \ll \tau_E$, is difficult to reconcile with a picture in which
momentum relaxation is intrinsically inelastic. It is, however, naturally
explained if momentum is transferred predominantly through elastic or
quasielastic processes, while energy relaxation proceeds through slower
coupling to internal lattice degrees of freedom.
\subsection{Shubnikov-de Haas oscillations}

Quantum oscillation phenomena such as the Shubnikov-de Haas and
de Haas-van Alphen effects provide a precise probe of electronic
coherence and scattering in metals at low temperature~\cite{Shoenberg1984,Ando1982}.
These oscillations arise from Landau quantization and require that
electrons execute many cyclotron orbits without loss of phase coherence.

Within the conventional Lifshitz-Kosevich (LK) framework, the oscillation
amplitude may be written as
\begin{equation}
A(T,B) \propto 
\frac{X}{\sinh X}
\exp\!\left(-\frac{\pi}{\omega_c \tau_q}\right),
\qquad
X=\frac{2\pi^2 k_B T}{\hbar \omega_c},
\end{equation}
where the factor $X/\sinh X$ describes thermal smearing of the Fermi
surface, while the Dingle factor
\begin{equation}
\exp\!\left(-\frac{\pi}{\omega_c \tau_q}\right)
\end{equation}
encodes Landau-level broadening through the quantum lifetime $\tau_q$.
In standard treatments, $\tau_q$ is associated with inelastic or
quasi-inelastic processes, and is often assumed to become weakly
temperature dependent at low $T$.

From the present viewpoint, this interpretation is unnecessarily
restrictive.  When the lattice center-of-mass degree of freedom is
retained, momentum relaxation need not proceed through phonon creation
or annihilation.  Instead, the electron can exchange momentum elastically
with the lattice background via recoil of the zero mode, while total
mechanical momentum is conserved.  These processes are elastic in energy,
but nonetheless randomize the electronic phase.

As a result, the Dingle factor should be understood more generally as
encoding phase randomization rather than strictly inelastic scattering.
Elastic recoil processes provide a continuous channel for Landau-level
broadening, even in the absence of phonon excitation.  The temperature
dependence then arises not from phonon occupation factors, but from the
growth of available recoil phase space in the thermally fluctuating
background.

This leads naturally to a temperature-dependent quantum lifetime of the form
\begin{equation}
\frac{1}{\tau_q(T)} \sim T^\alpha,
\end{equation}
with $\alpha \sim 1$ in a Planckian regime, implying
\begin{equation}
\exp\!\left(-\frac{\pi}{\omega_c \tau_q(T)}\right)
\;\sim\;
\exp\!\left(-\mathrm{const}\,\frac{T}{B}\right).
\end{equation}
Thus, the damping of quantum oscillations can remain strongly
temperature dependent down to low $T$ without invoking inelastic
scattering.

The persistence of Shubnikov-de Haas oscillations over wide temperature
ranges therefore indicates that substantial momentum randomization can
occur through predominantly elastic processes, while phase coherence is
degraded only gradually.  This resolves the apparent tension between
strong scattering and long-lived cyclotron coherence, and places quantum
oscillation damping on the same footing as weak localization: both probe
phase randomization in a dynamically fluctuating, but not necessarily
energy-relaxing, environment.

This interpretation is consistent with experimental observations of a
marked temperature dependence of the Dingle factor, often extending to
low $T$ with approximately power-law or linear-in-$T$ behavior, rather
than saturation.  In contrast to conventional theory, which attributes
Landau-level broadening primarily to inelastic processes and therefore
expects weak temperature dependence at low $T$, the present framework
predicts a continued $T$-dependent damping arising from elastic
recoil-driven phase randomization.




\section{Discussion}

Here we examine the consistency of the present framework with
well-established experimental results, including the Wiedemann–Franz
law, weak localization, and quantum oscillations.

Second, we outline broader interpretive implications for Planckian
transport, linear-in-$T$ resistivity, and the Mott–Ioffe–Regel crossover.
The latter discussion is exploratory and is presented as a coherent
physical interpretation rather than as a definitive claim.

\subsection{Implications for the Lorenz ratio}

Elastic scattering randomizes momentum as effectively as inelastic
scattering within standard transport formalisms.
Accordingly, a predominantly elastic microscopic channel does not
conflict with the established success of conventional transport theory.

In a degenerate metal, the Wiedemann–Franz (WF) law states that the
ratio of the electronic thermal conductivity $\kappa$ to the electrical
conductivity $\sigma$ times temperature approaches the Sommerfeld value
\begin{equation}
L \equiv \frac{\kappa}{\sigma T}
\;\xrightarrow[T\to 0]{}\;
L_0 = \frac{\pi^2}{3}\left(\frac{k_B}{e}\right)^2 .
\end{equation}
Although some inelastic processes must ultimately relax the electronic
energy current and establish global thermal equilibrium, it has long
been understood that these processes need not determine the leading
transport coefficients themselves~\cite{Ziman,Langenfeld1991,Reizer1986}.

If the dominant momentum-relaxing channel is elastic and only weakly
energy dependent over the thermal window $\sim k_B T$ about the Fermi
level, then both electrical and thermal currents are governed by the
same relaxation time $\tau_{\mathrm{el}}$,
\begin{equation}
\sigma \propto \tau_{\mathrm{el}}, \qquad
\kappa \propto \tau_{\mathrm{el}},
\end{equation}
and the Sommerfeld value $L_0$ follows in the usual way.
Inelastic processes with rate $1/\tau_{\mathrm{in}} \ll 1/\tau_{\mathrm{el}}$
transfer energy between electrons and lattice degrees of freedom but do
not enter the leading expressions for $\sigma$ or $\kappa$.
A transport regime dominated by elastic momentum randomization is
therefore fully compatible with the WF law, even when true energy
relaxation is parametrically slower.

\subsection{Weak localization}

Weak localization (WL) provides a clear illustration of the central
role of elastic backscattering in low-temperature metallic transport.
The phenomenon arises from constructive interference between pairs of
time-reversed electronic trajectories that revisit the same spatial
region after multiple scattering events~\cite{Bergmann1984,LeeRamakrishnan1985}.
It is often the case that the WL correction is important only with defects causing elastic backscattering.   However the electrons still have to arrive back intact, so to speak (i.e. in phase), so other inelastic events cannot have decohered the electron on the way out and back.  This is a tall order if any ``natural'' deflections are inelastic. elcrtherefore requires that a substantial fraction of
backscattering processes preserve phase coherence over the relevant
timescale.

If scattering were strongly inelastic on the transport timescale,
interference between time-reversed paths would be suppressed before
a WL correction could develop.
Experimentally, however, a well-defined suppression of conductivity
and characteristic negative magnetoresistance are observed at low
temperature.
As temperature increases, the WL correction is gradually reduced.
Although this reduction is commonly described in terms of increasing
``inelastic scattering,'' what is operationally measured is the
loss of phase coherence rather than a rapid increase in electronic
energy relaxation.

Independent probes, including hot-electron relaxation and Johnson-noise
thermometry, indicate that electron–phonon energy exchange can remain
comparatively slow in the same temperature range.
The suppression of WL is therefore consistent with the onset of a
time-dependent scattering environment that randomizes phase while
involving only modest energy transfer on electronic scales.

Within the present framework this behavior fits naturally into the
hierarchy of timescales
\[
\tau_{p} \ll \tau_{\phi} \ll \tau_{E},
\]
where elastic momentum randomization produces diffusive transport,
while slower temporal evolution of the lattice background reduces
phase coherence without requiring substantial phonon excitation
or rapid energy dissipation.

\subsection{Planckian diffusion and linear-in-$T$ resistivity}

Much recent literature discusses Planckian behavior in terms of a
microscopic relaxation time
\[
\tau_{\mathrm{Pl}} \sim \frac{\hbar}{k_B T},
\]
often interpreted as setting a characteristic scale for scattering
or energy dissipation~\cite{Zaanen2004,Hartnoll2015,HartnollMackenzie2022,Sachdev2011}.
Within this perspective, transport coefficients are commonly estimated
by inserting $\tau_{\mathrm{Pl}}$ into kinetic or hydrodynamic expressions.

The framework developed in Ref.~\cite{aydinNAS1,zhang_planckian_2024} and extended
here suggests a complementary interpretation.
The primary microscopic result is the emergence of real-space diffusion
with diffusion constant
\begin{equation}
D \sim \frac{\hbar}{m^\ast},
\end{equation}
arising from elastic momentum scrambling in a time-dependent lattice
background.
In this formulation, diffusion emerges from quantum dynamics in a
fluctuating potential and does not rely explicitly on strong inelastic
scattering or rapid energy dissipation.
Wave-on-wave simulations of the coupled electron–lattice problem
demonstrate this mechanism explicitly
\cite{aydinNAS1,PhysRevLett.132.186303,zhang_planckian_2024}.

Once diffusion is established, Einstein relations connect $D$ to
transport coefficients,
\begin{equation}
\sigma = \chi D, \qquad
D = \mu \frac{k_B T}{e},
\end{equation}
so that expressing $D$ in terms of an effective relaxation time yields~\cite{aydinNAS1}
\begin{equation}
\tau_{\mathrm{eff}} \sim \frac{\hbar}{k_B T},
\end{equation}
up to factors of order unity.
From this viewpoint, the Planckian timescale may be understood not as
a fundamental bound on inelastic scattering, but as an emergent
parametrization of quantum-limited diffusion combined with equilibrium
thermodynamics.
It characterizes the rate at which charge spreads spatially rather
than the rate at which energy must be irreversibly transferred to
internal degrees of freedom, and thus fits nicely into the main elastic and quasielastic theme of this paper.

    \subsection{Coulomb drag}
    We began discussing Coulomb drag in section~\ref{ring}. It is an ideal case in which the zero recoil mode, missing in the literature on coulomb drag, is likely the agent of the power law seen in experiments. Theoretical analyses of one-dimensional transport consistently
    emphasize the strong kinematic constraints imposed by simultaneous
    energy and momentum conservation, leading to the conclusion that
    relaxation processes should be strongly suppressed at low temperature
    \cite{Lunde2007ThreeParticle,Karzig2010HotElectrons} .
    Experimentally, however, relaxation and drag in quantum wires are
    often observed to exhibit power-law temperature dependence rather
    than strong suppression. This discrepancy is typically attributed to
    additional mechanisms such as disorder, band curvature, or coupling
    to external degrees of freedom that relax the kinematic constraints. This will be the topic of a future publication. 

  \subsection{Phonon drag}

In conventional transport theory, the phonon-drag contribution to the
thermopower is attributed to momentum transfer from a nonequilibrium
phonon population to the electrons, and is therefore tied to phonon
creation and annihilation. This reflects a formulation in which the
lattice center-of-mass (zero mode) is removed and momentum exchange is
forced into the phonon sector.

When the zero mode is retained, momentum transfer is no longer tied to
phonon number. The electron couples to the lattice deformation field,
and momentum can be exchanged with the lattice as a whole through
recoil, without requiring phonon excitation. The conventional
phonon-drag contribution therefore represents only a subset of a more
general lattice-drag response.

Accordingly, the thermopower should be written as
\begin{equation}
S = S_{\rm diff} + S_{\rm lat}, \qquad
S_{\rm lat} = S_{\rm el} + S_{\rm inel},
\end{equation}
where $S_{\rm inel}$ is the usual phonon-drag term and $S_{\rm el}$
arises from elastic or quasielastic momentum transfer via lattice
recoil. In regimes where scattering is predominantly elastic,
$S_{\rm el}$ need not be small and may dominate the drag response.

\section{Summary}

We have revisited electron-lattice scattering starting from the
standard microscopic Hamiltonian, but retaining a degree of freedom
that is nearly always removed: the translational center-of-mass (zero
mode) of the lattice. In conventional formulations, this mode is
eliminated (e.g., through Born-von K\'arm\'an boundary conditions),
leaving momentum transfer to be carried entirely by internal phonon
excitations. Together with the form of the first order Fr\"ohlich expansion, changes in electronic momentum are tied to
phonon creation or annihilation, and scattering is treated as
intrinsically inelastic.

When the lattice center of mass is retained explicitly, and the density-density form of the interaction is retained and not expanded, this picture
changes qualitatively. Total mechanical momentum is conserved by recoil
of the lattice as a whole, while pseudomomentum is redistributed between
electron and background. A broad class of elastic and quasielastic
momentum-transfer processes then becomes available, in which electronic
momentum changes are balanced by the zero mode without requiring phonon
excitation.

This enlargement of the allowed phase space follows directly from the
exact symmetries of the Hamiltonian. No new interaction has been
introduced. Retaining the full density-density interaction reveals that
elastic matrix elements arise already at the level of the state-to-state
transition amplitude $\langle f|H_{\rm int}|i\rangle$, without invoking
higher-order or multiphonon processes. The familiar phonon-mediated
processes remain, but represent a restricted subset of a much larger set
of allowed transitions.

The resulting kinematics has broad implications. In one-dimensional
systems, the need for thermally activated three-particle processes is
removed, and power-law relaxation follows naturally. More generally,
momentum relaxation need not be tied to energy dissipation: momentum can
randomize rapidly while energy relaxes on much longer timescales. This
resolves a long-standing tension between theory and experiment in clean
metals, including observations from weak localization, quantum
oscillations, hot-electron transport, and Coulomb drag.

We emphasize that these conclusions are not model-dependent. Once the
full momentum conservation law of the isolated electron-lattice system
is enforced, the existence of elastic and quasielastic scattering
channels is unavoidable. The restriction to phonon-mediated inelastic
scattering is therefore not a fundamental feature of the physics, but a
consequence of having discarded the lattice center-of-mass degree of
freedom.

\section{Acknowledgments}

E.~Heller thanks B.~Halperin for many discussions that provided a thorough understanding of the traditional theory. He also thanks Prof.~Martina~Hentschel for insightful comments that have improved the manuscript. This work was supported by the U.S.\ Department of Energy, Office of Science, under Grant No.\ DE-SC0025489.

\begin{figure}
        \centering
\includegraphics[width=0.85\linewidth]{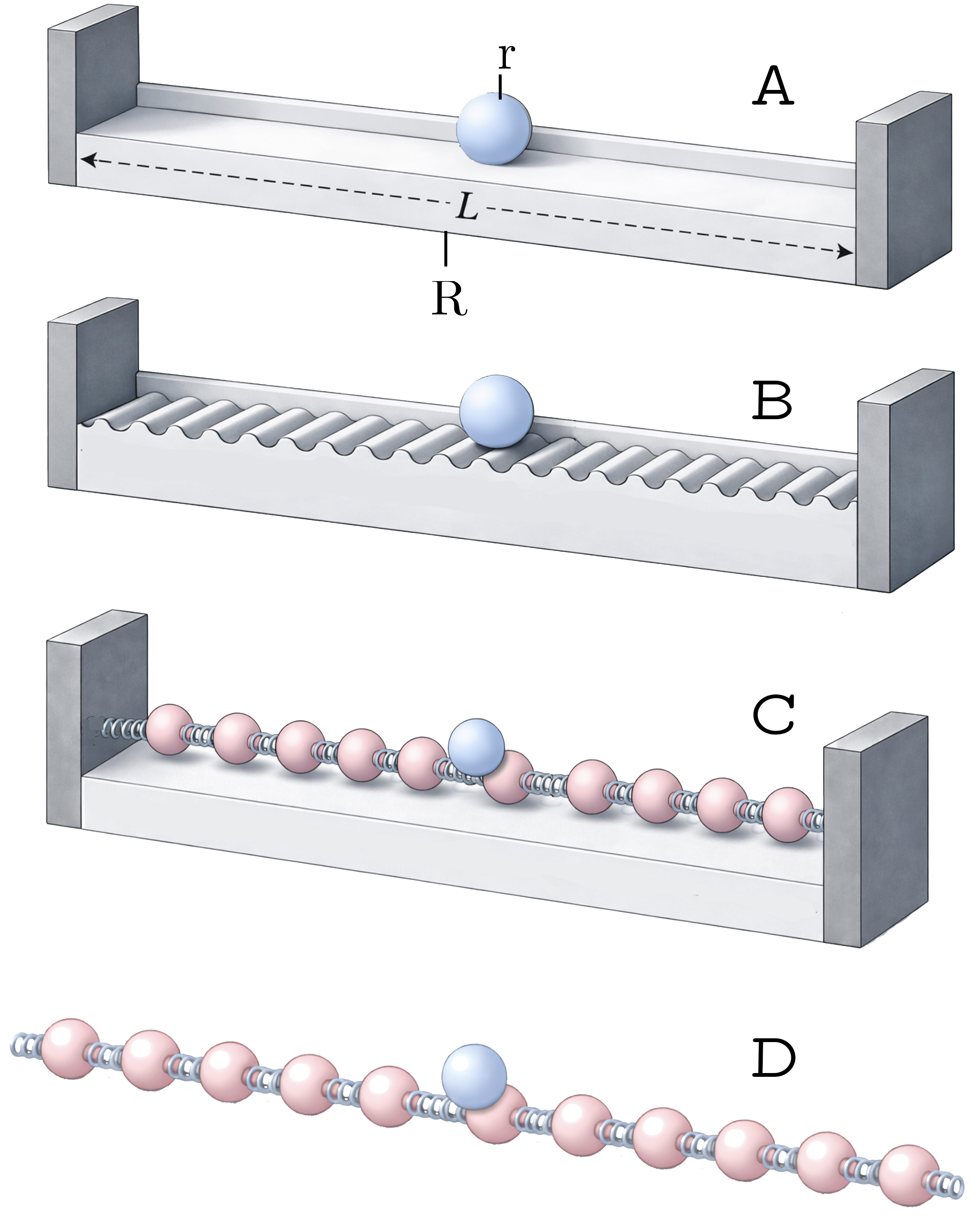}
    \caption{Isolated systems of mass $M+m$ containing a foreground particle of mass $m$ and a background of mass $M$.  The particle is attracted to the background and moves frictionlessly, except in C and D it may excite vibrations of the chain. In A, the particle is free except for walls. In B, the particle is in a fixed periodic potential within its host, the box. In C we replace the fixed corrugation with a chain of masses connected by springs. In D, we dispense with the box, coming close to the situation that matters here. Of course the box could be sitting on a  table, which could be fixed to a floating laboratory, which could be upon something so massive that it has significant gravitational attraction, yet that too is  floating in space...    }
\label{fig:cases}
        \end{figure}
               

\appendix

\section{Bloch waves form from elastic scattering of electrons in a lattice}
By way or reinforcing the conclusion that elastic internal electron scattering must be very common in pure metals, we raise a point about the rather violent elastic scattering that is taking place, cl    oacked by the symmetry of Bloch waves. The formation of Bloch waves in a periodic crystal may be viewed as the result of coherent multiple elastic scattering from the atoms of the lattice\cite{Kohn1964}. Band formation itself is therefore an elastic scattering phenomenon. One might then suppose that Bloch waves are somehow protected from further elastic deflection by phase coherence. However, this inference is incorrect, as illustrated in Fig.~\ref{Kohnspoint}.

 We extend this insight slighty and suppose some atom were temporarily slightly out of place due to thermal acoustic waves. The atoms would individually scatter as before, but now the phase locking is imperfect, and the electron can deflect.  There is no new ``phonon needed''  physics that sets in due to truly minuscule thermal displacements. If the electron thus deflects, momentum  is
exchanged with the lattice without phonon excitation or energy
dissipation.
\section{ Tracking Momentum, and the Effects of Zero Mode Removal}

Figure~\ref{fig:cases} helps illuminate the choices that
can be made in treating   electron-phonon interaction.  It helps   to underline that there is a great difference between treating the background
 as a dynamical body (with a translational zero mode) or as  frozen, without  recoil.  Frozen is an impossibility in the real world, total mometum is always conserved,  although this a self-consistent model. The question is, does that model lead to good real-world predictions, or does it mislead? Retaining the zero mode total momentum  is not a matter of interpretation; it fixes the
 form of the wavefunction and alters the physical predictions in important ways. 

Translation invariance of the whole requires that the
Hamiltonian depends only on the center of mass and coordinate differences, so eigenstates separate
into a plane wave for the center of mass coordinate and an internal state that
depends only on   \emph{relative} coordinates.

\subsection{Uplifting the particle in   a box to the real world.}
We therefore free the textbook particle in a box from its traditional and physically impossible bondage in case A.: considering a``free floating box'' of mass $M$ containing a particle of mass $m$,
with particle coordinate $r$ and box center-of-mass coordinate $R$, introduce
\[
\mathcal R=\frac{mr+MR}{m+M},\qquad 
\]
Then a complete set of eigenstates (delta-normalized in $P$) may be written as

\begin{equation}
\Psi_{P,n}(r,R)
=
\frac{1}{\sqrt{2\pi\hbar}}\,
e^{iP\mathcal R/\hbar}\,
\phi_n(r-R).
\label{eq:rR_wavefunction}
\end{equation}
Equation~\eqref{eq:rR_wavefunction} is the operative content of keeping the
zero mode: the internal structure is \emph{literally} a function of $r-R$.
The conserved total mechanical momentum is
\[
\hat P_{\rm tot}=-i\hbar\frac{\partial}{\partial\mathcal R},
\qquad
[H,\hat P_{\rm tot}]=0.
\]
Any internal state may be expanded in Fourier components on $|r-R|<L/2$,
\[
\phi_n(r-R)=\sum_q a_{n q}\,e^{iq(r-R)},
\]
and each component carries the shared phase
\begin{equation}
e^{iq(r-R)}=e^{iqr}e^{-iqR}.
\label{eq:shared_phase_here}
\end{equation}
Thus, in the plane-wave decomposition of the relative coordinate, momentum
exchange is enforced \emph{at the level of the wavefunction}: each Fourier
component correlates $+\hbar q$ on the particle with $-\hbar q$ on the
background.  Phonons and other internal excitations may redistribute
momentum within the background.

\subsection{Box is corrugated} 

Suppose the box is as above but now contains a rigid, corrugated
periodic potential (Fig.~\ref{fig:cases}, B). Before the corrugation,
the particle and the box each carry mechanical momentum, and only the
total
\(
P_{\rm tot} = p + P
\)
is conserved. In collisions, momentum is exchanged between the two in
equal and opposite amounts, exactly as in free space.

The introduction of a periodic potential does not eliminate this
structure, but reorganizes it. Continuous translational symmetry is
reduced to discrete translational symmetry, and the corresponding
conserved quantities become pseudomomenta defined modulo reciprocal
lattice vectors. Importantly, the two-body structure of momentum
exchange is preserved.

The particle now carries a pseudomomentum $\hbar k$, while the
background box carries a corresponding pseudomomentum. These two
pseudomomenta are exchanged in equal and opposite amounts during
scattering, just as the mechanical momenta were exchanged before the
corrugation was introduced. Their sum remains fixed and corresponds to
the conserved total momentum of the isolated system.

Thus the corrugation does not replace two mechanical momenta with a single
pseudomomentum; rather, it produces two pseudomomenta, one for the
particle and one for the background. The center-of-mass degree of
freedom of the box continues to carry the total mechanical momentum,
while the periodic structure introduces a new, discrete labeling of the
same underlying exchange process.

The situation depicted in Fig.~\ref{fig:cases}, B therefore retains the
essential physics of momentum exchange: the lattice remains a dynamical
participant, even though its role is partially obscured by the use of
pseudomomentum.

\subsection{Elastic Chain in a Box, Free CM (phonons \emph{plus} the zero mode)}

Now take the background to be a dynamical chain (pink atoms) of total
mass $M_{\rm lat}$ inside the box, with atomic coordinates $R_j$ and a
center-of-mass coordinate $R_{\rm lat}$ (Fig.~\ref{fig:cases}, C). A
minimal model is
\[
H=
\frac{p_r^2}{2m}
+
\sum_j\frac{p_j^2}{2M}
+
\frac{K}{2}\sum_j(R_{j+1}-R_j-a)^2
+
H_{\rm int}.
\]
With density-density coupling
\[
H_{\rm int}=
\int dr\,dr'\,\rho_p(r)V(r-r')\rho_{\rm lat}(r'),
\]
the Hamiltonian depends only on coordinate differences and is invariant
under rigid translation
\[
r\rightarrow r+a_0,\qquad R_j\rightarrow R_j+a_0.
\]
Consequently the conserved quantity is the true total mechanical momentum
\[
P_{\rm tot}=P_e+P_{\rm lat},\qquad [H,P_{\rm tot}]=0.
\]

Momentum exchange is enforced directly through the relative coordinate.
Fourier transforming the lattice density yields
\[
\rho_{\rm lat}(q)=e^{-iqR_{\rm lat}}\rho^{\rm int}_q,
\qquad
H_{\rm int}\propto e^{iq(r-R_{\rm lat})}.
\]
The phase factor $e^{iq(r-R_{\rm lat})}$ encodes the fundamental
constraint: momentum $\hbar q$ transferred to the electron is accompanied
by $-\hbar q$ carried by the lattice center of mass. Internal degrees of
freedom describe only relative motion and carry no net mechanical
momentum.

The translational zero mode $R_{\rm lat}$ is therefore the unique carrier
of the compensating momentum. Internal modes (phonons) may change the
energy of the system, but they are not required to enforce momentum
conservation.

Deleting the zero mode is not an innocuous simplification. It removes the
only degree of freedom capable of carrying recoil and eliminates the
shared-coordinate structure $\phi(r-R)$ from the Hilbert space. The
resulting theory no longer represents true mechanical momentum
conservation; instead it replaces it with pseudomomentum bookkeeping
within a reduced space. Momentum exchange is then no longer enforced
through the relative coordinate $r-R$, but appears only as a selection
rule requiring phonon pseudomomentum to balance electron pseudomomentum
(up to Umklapp).

Thus the apparent kinematic necessity of phonon production does not arise
from fundamental principles. It is imposed by amputating the translational
degree of freedom that would otherwise carry the compensating momentum.

\subsection{Free Chain (no box), Free CM (translation invariance made manifest)}

Removing the box entirely leaves a floating chain. This is simply
Fig.~\ref{fig:cases}, C with boundary constraints eliminated. The lesson
is unchanged but becomes explicit: the conserved momentum is the
generator of rigid translation of \emph{all} coordinates.

Any description that retains translational invariance necessarily
contains the shared phase $e^{iq(r-R_{\rm lat})}$, since it is the
Fourier representation of the relative coordinate. This phase structure
is the direct expression of momentum conservation in the full system.

\subsection{Removing the Zero Mode}

The conventional electron-phonon kinematics is obtained by a
construction that is the Hilbert-space analog of ``nailing the box
down,'' but without introducing an external dynamical reservoir. The
translational zero mode is removed at the outset. One expands the lattice
displacement field in normal modes and omits the $q=0$ component. The
resulting description retains only internal degrees of freedom.

With the zero mode removed, continuous translation symmetry is no longer
present, and the conserved quantity is not the true mechanical momentum.
What remains is discrete lattice translation symmetry, with conserved
crystal pseudomomentum
\[
K_{\rm tot}=K_e+K_{\rm ph}\quad(\mathrm{mod}\;G),
\qquad
K_{\rm ph}=\sum_q\hbar q\,a_q^\dagger a_q,
\]
and the corresponding selection rule
\[
\Delta K_e+\Delta K_{\rm ph}=Gm.
\]

In this representation, momentum transfer is confined to  phonons: changes in the electron pseudomomentum must be balanced entirely
by phonon pseudomomentum, up to Umklapp. The lattice center-of-mass
degree of freedom, which would otherwise provide a direct channel for
momentum exchange through rigid recoil, is absent.

The apparent kinematic necessity of phonon emission or absorption is
therefore not a physical result. It is a  consequence of
having eliminated the translational degree of freedom that would
otherwise carry the compensating momentum. It is not a choice one normally would wish to make, because it changes the predicted physical processes. It is surprising perhaps, but it does. The ``thermodynamic limit'' does not protect us here, and if it seems that it does, the limit $N\to \infty$ has not been taken properly. Jumping to $N= \infty$ is not taking the limit of course. 
\section{Whither Umklapp?}
\label{app:whither_umklapp}

The thermodynamic limit and Umklapp processes are often invoked to
account for momentum relaxation in perfect crystals. These mechanisms are
valid within the conventional formulation, but their apparent necessity
reflects the prior removal of the lattice center-of-mass (zero-mode)
degree of freedom.

In the standard phonon description, momentum conservation is imposed only
at the level of crystal pseudomomentum,
\[
K_{\rm tot}=K_e+K_{\rm ph}\quad(\mathrm{mod}\;G),
\]
and electronic momentum can relax only through phonon processes,
supplemented by Umklapp when reciprocal lattice vectors are invoked. This
framework is kinematically restrictive: the phonon system tracks momentum
only modulo $G$, and the available channels are limited to discrete
reciprocal lattice transfers.

When the zero mode is retained, this restriction is lifted. The lattice
background can recoil and absorb arbitrary mechanical momentum, and the
scattering problem is no longer confined to Umklapp processes. Instead,
momentum transfer occurs within a continuous manifold of elastic and
quasielastic processes involving the full dynamical background.

To make this explicit, we separate a single electron (the foreground)
from the rest of the system (the background), which includes all ions and
all other electrons. The total mechanical momentum is
\begin{equation}
\hat P_{\rm tot}
=
\hat p
+
\hat P_{\rm bg}, \qquad
\hat P_{\rm bg}=\sum_{\alpha \in \mathrm{bg}} \hat P_\alpha.
\end{equation}
The operator $\hat P_{\rm bg}$ generates rigid translation of the
background and constitutes its center-of-mass (zero-mode) momentum.
Internal phonon modes describe only relative motion and do not furnish an
independent contribution to $\hat P_{\rm bg}$. Thus any change in the
electron’s mechanical momentum must be balanced by the background zero
mode.

This makes clear what is implicit in Umklapp. When a reciprocal lattice
vector $G$ appears in a scattering process, the associated momentum
$\hbar G$ cannot be attributed to the electron alone, nor is it carried by
internal phonon modes. It must be carried by the lattice as a whole. In
the conventional formulation, however, the zero mode that would carry this
momentum has been removed, and the role of recoil is replaced by
bookkeeping modulo $G$.

The traditional role assigned to Umklapp in producing finite resistivity
therefore arises within a restricted description in which the background
is effectively clamped. Once the background is treated dynamically,
electronic momentum need not decay through Umklapp. Instead, it can be
transferred elastically to the lattice as a whole.

Umklapp remains a useful classification of pseudomomentum conservation
modulo reciprocal lattice vectors. It is not, however, the unique or even
generic mechanism for momentum relaxation in a real crystal, where the
background is free to recoil.

\section{The landscape seen by an electron}

At temperatures of order $T\sim 100\,$K, thermally excited acoustic distortions in
a typical metal generate substantial deformation-potential landscapes on
nanometer length scales.
Using standard values for the deformation potential
($g\sim 5$–$15\,\mathrm{eV}$), sound velocity
($v_s\sim 3$–$5\times10^3\,\mathrm{m/s}$), and the characteristic thermal phonon
wavevector $q_T\sim k_B T/\hbar v_s$, one finds order-of-magnitude estimates for
the resulting deformation-potential field gradients of
\begin{equation}
|\nabla V_{\rm def}|
\sim
10^{4}\text{–}10^{6}\ \mathrm{V/cm}
\qquad (T\sim 100\,\mathrm K).
\end{equation}
Over a representative lateral scale of $L\sim 10\,$nm, this corresponds to
peak-to-peak variations of the screened scalar potential of order
\begin{equation}
\Delta V_{\rm pp}
\sim
|\nabla V_{\rm def}|\,L
\sim
10\text{–}300\ \mathrm{mV}.
\end{equation}

A deformation-potential landscape of order
\(0.1\)–\(0.3\,\mathrm{eV}\) can have very different physical consequences in
conventional metals and in strange metals.
In a metal such as Cu, the Fermi energy and bandwidth are large
(\(E_F \sim 7\,\mathrm{eV}\)), electronic states near the Fermi surface are
long-lived and coherent, and electronic screening enforces local charge
neutrality on fast timescales.
As a result, lattice-induced scalar potentials of this magnitude—already
understood as screened, low-energy deformation potentials—enter transport only
as weak, slowly varying perturbations.
The resulting density response is small, quasiparticles remain well defined, and
transport is accurately described by perturbative scattering theories such as
Bloch-Gr\"uneisen~\cite{Ziman}.

The situation is qualitatively different in optimally doped strange metals.
There, the effective Fermi energy and bandwidth are much smaller, the electronic
compressibility is large and strongly temperature dependent, and electronic
states near the Fermi level cannot be described in terms of long-lived Bloch
quasiparticles.
In this regime, a deformation potential of order
\(0.1\)–\(0.3\,\mathrm{eV}\) is comparable to the relevant electronic energy
scales and cannot be treated as a weak perturbation.
Instead, it drives substantial rearrangements of the local electronic density,
strongly couples the electronic fluid to the dynamical lattice background, and
precludes any residual adiabatic separation between electronic and lattice
degrees of freedom.
When such a potential varies in both space and time, it does not produce static
localization but instead leads to diffusive electronic motion, naturally giving
rise to Planckian-scale transport with a diffusion constant of order
\(D \sim \hbar/m\)
\cite{aydinNAS1,zhang_planckian_2024,heller_quantum_2025}.

\bigskip

\bibliographystyle{apsrev4-2}
\bibliography{apssamp}

\end{document}